\documentclass[12pt]{article}
\pdfoutput=1
\usepackage{jheppub}
\usepackage{amsmath}
\usepackage{amsfonts}
\usepackage{amssymb}
\usepackage{graphicx}
\usepackage[export]{adjustbox}
\newcommand{\bmat}{\left(\begin{array}}
\newcommand{\emat}{\end{array}\right)}

\def\yzero{\smash{\hbox{$y\kern-4pt\raise1pt\hbox{${}^\circ$}$}}}

\def\beq{\begin{equation}}
\def\eeq{\end{equation}}
\def\beqa{\begin{eqnarray}}
\def\eeqa{\end{eqnarray}}

\def\-{\hphantom{-}}
\def\ov{\overline}
\def\s2{\frac{1}{\sqrt2}}

\def\beq{\begin{equation}}
\def\eeq{\end{equation}}
\def\beqa{\begin{eqnarray}}
\def\eeqa{\end{eqnarray}}
\def\tr{{\rm tr \,}}

\def\II{\relax{\rm I\kern-.18em I}}

\def\Dsl{\,\raise.15ex\hbox{/}\mkern-13.5mu D} %this one can be subscripted
\def\IC{{\bf{C}}}
\def\IF{{\bf{F}}}
\def\IS{{\bf {S}}}
\def\IR{{\bf {R}}}
\def\IZ{{\bf {Z}}}
\def\IX{{\bf {X}}}
\def\IY{{\bf {Y}}}

\def\IP{{\bf {P}}}

\def\NN{{\cal {N}}}

\newcommand{\drawsquare}[2]{\hbox{%
\rule{#2pt}{#1pt}\hskip-#2pt%  left vertical
\rule{#1pt}{#2pt}\hskip-#1pt%  lower horizontal
\rule[#1pt]{#1pt}{#2pt}}\rule[#1pt]{#2pt}{#2pt}\hskip-#2pt%  upper horizontal
\rule{#2pt}{#1pt}}% right vertical

\newcommand{\fund}{\,\raisebox{-.5pt}{\drawsquare{6.5}{0.4}}\,}%  fund
\newcommand{\asymm}{\,\raisebox{-3.5pt}{\drawsquare{6.5}{0.4}}\hskip-6.9pt%
        \raisebox{3pt}{\drawsquare{6.5}{0.4}}\,}%  antisymmetric second rank
\newcommand{\antifund}{\overline{\fund}}

\catcode`\@=11   
\newdimen\@rotdimen
\newbox\@rotbox  

\def\@vspec#1{\special{ps:#1}}%  passes #1 verbatim to the output
\def\@rotstart#1{\@vspec{gsave currentpoint currentpoint translate
   #1 neg exch neg exch translate}}% #1 can be any origin-fixing transformation
\def\@rotfinish{\@vspec{currentpoint grestore moveto}}% gets back in synch 
\def\@rotr#1{\@rotdimen=\ht#1\advance\@rotdimen by\dp#1%
   \hbox to\@rotdimen{\hskip\ht#1\vbox to\wd#1{\@rotstart{90 rotate}%
   \box#1\vss}\hss}\@rotfinish}
\def\@rotl#1{\@rotdimen=\ht#1\advance\@rotdimen by\dp#1%
   \hbox to\@rotdimen{\vbox to\wd#1{\vskip\wd#1\@rotstart{270 rotate}%
   \box#1\vss}\hss}\@rotfinish}%
\def\@rotu#1{\@rotdimen=\ht#1\advance\@rotdimen by\dp#1%
   \hbox to\wd#1{\hskip\wd#1\vbox to\@rotdimen{\vskip\@rotdimen
   \@rotstart{-1 dup scale}\box#1\vss}\hss}\@rotfinish}%
\def\@rotf#1{\hbox to\wd#1{\hskip\wd#1\@rotstart{-1 1 scale}%
   \box#1\hss}\@rotfinish}%
\def\rotate{\@ifnextchar[{\@rotate}{\@rotate[l]}}
\def\@rotate[#1]#2{\setbox\@rotbox=\hbox{#2}\@nameuse{@rot#1}\@rotbox}

\catcode`\@=12
\begin{document}

%----------------------------------------------------------------------%
%  numbering equations with section number
%----------------------------------------------------------------------%
\makeatletter
\@addtoreset{equation}{section}
\makeatother
\renewcommand{\theequation}{\thesection.\arabic{equation}}
%----------------------------------------------------------------------%
%  title page
%----------------------------------------------------------------------%
\pagestyle{empty}
\vspace*{0.5in}
\rightline{IFT-UAM/CSIC-24-136}
\vspace{1.5cm}
\begin{center}
\Large{\bf End of the World Boundaries \\for Chiral Quantum Gravity Theories 
}
\\[8mm] 
%}\\

\large{Roberta Angius, Angel M. Uranga, Chuying Wang \\[4mm]}
\footnotesize{Instituto de F\'{\i}sica Te\'orica IFT-UAM/CSIC,\\[-0.3em] 
C/ Nicol\'as Cabrera 13-15, 
Campus de Cantoblanco, 28049 Madrid, Spain}\\ 
\footnotesize{\href{roberta.angius@csic.es}{roberta.angius@csic.es},  \href{mailto:angel.uranga@csic.es}{angel.uranga@csic.es}, \href{mailto:chuying.wang@ift.csic.es}{chuying.wang@ift.csic.es}}

\vspace*{10mm}

\small{\bf Abstract} \\%[5mm]
\end{center}
\begin{center}
\begin{minipage}[h]{\textwidth}
We describe the construction of large classes of explicit string theory backgrounds corresponding to 6d and 4d chiral theories with end of the world boundaries, and describe the strong coupling phenomena involved in gapping the chiral (but non-anomalous) sets of fields, such as strongly coupled phase transitions or symmetric mass generation. One class of 6d constructions is closely related to chirality changing phase transitions, such as those turning heterotic NS5-branes into gauge instantons, in flat space or orbifold singularities. A class of 4d models exploits systems of IIB D3-branes at toric CY3 singularities with an extra $\IZ_2$ involution related to $G_2$ holonomy manifolds in the type IIB picture and its IIA mirror, which we explicitly describe in terms of dimer diagrams.
\end{minipage}
\end{center}
\newpage
%----------------------------------------------------------------------%
%  Resetting of counters
%----------------------------------------------------------------------%
\setcounter{page}{1}
\pagestyle{plain}
\renewcommand{\thefootnote}{\arabic{footnote}}
\setcounter{footnote}{0}
%----------------------------------------------------------------------%
%  Paper begins
%----------------------------------------------------------------------%

\tableofcontents

\vspace*{1cm}

\newpage

\section{Introduction}
\label{sec:intro}

The swampland cobordism conjecture \cite{McNamara:2019rup} implies that any theory of quantum gravity must admit configurations including boundaries ending spacetime (end of the world or ETW configurations). These has been discussed in various contexts (see e.g. \cite{GarciaEtxebarria:2020xsr,Montero:2020icj,Dierigl:2020lai,Hamada:2021bbz,Buratti:2021fiv,Angius:2022aeq,Ooguri:2020sua,Dierigl:2023jdp,Debray:2023yrs}), but are particularly challenging for chiral theories. Indeed, even for string theory or M-theory in their maximal dimensions, such boundary configurations are essentially known only\footnote{One may also wish to include bosonic string theory and some supercritical string theories, for which analogues of bubbles of nothing have been built using light-like tachyon condensation \cite{Hellerman:2006nx,Hellerman:2006ff,Hellerman:2007fc}} for 11d M-theory (in the form of Ho\v{r}ava-Witten boundaries) and 10d type IIA (a negatively charged O8-plane with 16 D8-branes as counted in the double cover). This is intimately related to the fact that these theories are non-chiral at the level of their spectrum, and only break parity via topological Chern-Simons terms. In fact, for 10d type IIB, type I or heterotic theories, as well as their non-supersymmetric cousins, which are chiral yet anomaly free, there is no microscopic understanding of such boundary ETW configurations. Similar statements can be made in compactifications to lower dimensions.

Morally, the rationale for this relation is that in theories with vector-like spectrum the boundary conditions pair up opposite-chirality degrees of freedom. This is the equivalent of a gapping vector-like fermions with a Dirac mass. Thus, from this perspective, chirality prevents the existence of weakly coupled mechanisms to gap the set of chiral fermions, hence boundary conditions for chiral theories must involve strong coupling. This makes it difficult to formulate such boundary conditions, even in situations with high supersymmetry.

Although examples of mechanisms gapping chiral non-anomalous sets of fermions have been studied in the context of quantum field theory (see e.g. \cite{Razamat:2020kyf,Tong:2021phe}, also \cite{Wang:2022ucy} for a review), examples of boundary configurations for chiral theories in the context of quantum gravity or string theory are very scarce (one example is given by the bubble of nothing in \cite{Fabinger:2000jd}, when regarded from the 10d perspective; see also \cite{Friedrich:2023tid} for a proposed construction in 4d compactifications). In this paper we take important steps in improving this situation.

We build explicit boundary ETW configurations for large classes of examples of 6d and 4d chiral theories from string theory compactifications, hence coupled to quantum gravity. The examples are constructed by considering a $(d-1)$-dimensional locus of a $d$-dimensional localized chiral field theory in $D$-dimensional spacetime, and regarding the local configuration as a cone over the angular manifold in the $(D-d+1)$-dimensional transverse space around the $(d-1)$-dimensional slice. We are thus left with a compactification on the $(D-d)$-dimensional base of the cone, with a potentially chiral spectrum including the $d$-dimensional field theory. The cone defines a boundary configuration for the system, with an ETW boundary specified by the $(d-1)$-dimensional slice, which sits at the tip of the cone. The actual appearance of chirality in the $d$-dimensional theory is highly non-trivial and requires special physics happening at the $(d-1)$-dimensional locus, the tip of the cone. We dub this the Cone Construction or, when it leads to boundary configurations for actual chiral theories, the Chiral Cone Construction.

The Cone Construction provides an explicit link with the Dynamical Cobordisms of the compactified theory, in the sense of \cite{Buratti:2021fiv,Angius:2022aeq,Blumenhagen:2022mqw,Blumenhagen:2023abk}\footnote{For
  related ideas, see
  \cite{Dudas:2000ff,Blumenhagen:2000dc,Dudas:2002dg,Dudas:2004nd,Hellerman:2006nx,Hellerman:2006ff,Hellerman:2007fc}
  for early references, and
\cite{Basile:2018irz,Antonelli:2019nar,GarciaEtxebarria:2020xsr,Mininno:2020sdb,Basile:2020xwi,Mourad:2021qwf,Mourad:2021roa,Basile:2021mkd,Mourad:2022loy,Angius:2022mgh,Basile:2022ypo,Angius:2023xtu,Huertas:2023syg,Mourad:2023ppi,Angius:2023uqk,Delgado:2023uqk,Friedrich:2023tid,Angius:2024zjv,Mourad:2024dur,Mourad:2024mpg,GarciaEtxebarria:2024jfv,Ruiz:2024gzv}
  for recent works.}. In the Cone Construction, the lower-dimensional theory is obtained by compactification on the base of the cone. The evolution in the radial direction, along which the size of the compactification space varies, defines a solution with a running scalar for this lower dimensional theory. At the tip of the cone the corresponding scalar blows up to infinite field theory distance at a finite spacetime distance producing a singularity at which spacetime ends. This precisely agrees with the behaviour near an ETW configurations in Dynamical Cobordisms, and in particular there is a precise match with the local dynamical cobordism solutions in \cite{Angius:2022aeq} at the quantitative level.

Regarding the special physics at the $(d-1)$ slice, we specifically consider two main classes of models: 

$\bullet$ The first involves chirality changing phase transitions: We focus on explicit examples of 6d $\NN=1$ theories with heterotic NS5-branes reaching the origin of the Coulomb branch of their tensor multiplets and turning into gauge instantons, effectively trading each tensor multiplet for 29 hypermultiplets \cite{Seiberg:1996vs}. We consider several examples based on 5-branes in flat space or on orbifold singularities \cite{Seiberg:1996vs,Aldazabal:1996du,Aspinwall:1996vc,Aspinwall:1997ye,Intriligator:1997kq,Blum:1997mm,Blum:1997fw,Brunner:1997gf,Hanany:1997gh}, and apply the Cone Construction to obtain boundary configurations for large classes of 6d chiral theories.

$\bullet$ The second involves fixed planes under $\IZ_2$ involutions, closely related to those turning a CY3 conical singularity times $\IR$ into a (barely) $G_2$ holonomy variety \cite{Joyce:1996,Harvey:1999as}. We consider large classes of chiral 4d theories arising from IIB D3-branes at toric CY3 singularities, and use $\IZ_2$ quotients related to $G_2$ varieties in the IIA mirror, to define boundary conditions from Chiral Cone constructions. We exploit the powerful language of dimer diagrams as an efficient tool to describe the theories and the quotients leading to boundary configurations.

The paper is organized as follows. In section \ref{sec:transitions}, we consider explicit examples based on chirality changing phase transitions. After a warm-up in section \ref{sec:twodim} revisiting open heterotic strings (section \ref{sec:open-het}) and building cone construction over their boundaries (section \ref{sec:cone-open-het}), we move into the non-trivial case of Chiral Cone Constructions for 6d theories in section \ref{sec:sixdim}. We revisit the chirality changing phase transition for the $E_8\times E_8$ heterotic NS5-brane in flat space in section \ref{sec:M5}, and in section \ref{sec:open-ns5} we use the Cone Construction to define boundary configurations for chiral 6d theories. In section \ref{sec:open-m5} we relate our discussion to the supergravity solution \cite{Bergshoeff:2006bs} and its recent worldsheet description in \cite{Kaidi:2023tqo}. In section \ref{sec:5branes-at-singus} we extend our construction to 5-branes at singularities, and in section \ref{sec:symtft} we discuss relations with the cone constructions used in the string theory derivation of SymTFTs \cite{Apruzzi:2021nmk} in the study of generalized symmetries  (see \cite{McGreevy:2022oyu,Brennan:2023mmt,Gomes:2023ahz,Shao:2023gho,Schafer-Nameki:2023jdn,Bhardwaj:2023kri,Iqbal:2024pee} for reviews).

In section \ref{sec:fourdim} we focus on boundary configurations for 4d chiral theories. In section \ref{sec:open-intersecting} we emphasize how non-trivial the task is. We review intersecting D6-branes in section \ref{sec:inters} and open D6-branes ending on NS5 branes in section \ref{sec:opend6}, using them to construct localized 4d fermions on a space with boundary in section \ref{sec:4dferm-bdry}. However, in section \ref{sec:missing} we show that the corresponding Cone Construction fails (in an interesting way) to provide boundary conditions for chiral fermions, due to the presence of additional D4-branes. Overcoming this failure motivates the construction in section \ref{sec:cone-singus} of chiral gauge sectors localized on D3-branes at singularities, whose Cone Construction produces boundary configurations via a mechanism resembling that in  \cite{Fabinger:2000jd}. In section \ref{sec:dp0-example} we present one example leading to boundary conditions for the chiral 4d theory of D3-branes at $\IC^3/\IZ_3$ (the dP$_0$ theory), which in section \ref{sec:examples-singus} we extend to D3-branes at general CY3 toric singularities. In these models the special physics at the tip of the cone can be associated to brane-antibrane annihilation. In section \ref{sec:z2-quotients} we improve over this class of models, by including a $\IZ_2$ quotient ultimately lying at the tip of the cone. In section \ref{sec:g2} we motivate the construction by considering
the $\IZ_2$ quotients turning CY3$\times \IR$ into a barely $G_2$ holonomy variety. The mirror of such $\IZ_2$ actions is applied in section \ref{sec:mirror-d3s} to construct boundary configurations for theories arising from D3-branes at CY3 cone singularities, with several explicit examples described in section \ref{sec:examples}, and \ref{sec:fractional}. In section \ref{sec:dyn-cob-d3s} we describe the relation of the cone constructions with Dynamical Cobordisms. We study the general dimensional reduction in section \ref{sec:reduction}, particularize to compactification on the base of cones in section \ref{sec:base}, and show our cone constructions agree with the local dynamical cobordisms solutions in \cite{Angius:2022aeq} in section \ref{sec:scaling}.

In section \ref{sec:conclu} we offer some final remarks. In appendix \ref{app:more-intersecting} we extend the analysis of section \ref{sec:open-intersecting} to even more intricate configurations of intersecting D6-branes with boundaries, and show that their cone constructions do not lead to boundary conditions for 4d chiral theories. In appendix \ref{app:g2-cone-example} we revisit a system studied in \cite{Acharya:2003ii} and show it can be regarded as an explicit example of a $G_2$ cone construction providing boundary configuration for a 4d chiral gauge theory from intersecting D6-branes.

\section{Boundaries from Chirality changing phase transitions}
\label{sec:transitions}

The problem of gapping a set of chiral non-anomalous fields has appeared in string theory context in a slightly different avatar: the study of chirality changing phase transitions. In this section we argue that this question is closely related to the construction of boundary configurations for chiral theories via the Cone Construction, and present several classes of examples. 

\subsection{The Cone construction: Warm-up with the open heterotic string}
\label{sec:twodim}

In this section we introduce the key ideas of building boundary configurations for potentially chiral theories (the Chiral Cone construction), in terms of the example of the open heterotic string. The construction is easily generalized to other setups, as we study in later sections.

\subsubsection{Open heterotic string}
\label{sec:open-het}

A prominent manifestation of the difficulty to introduce boundary conditions for chiral theories arises in the context of D-branes as defining boundary conditions for 2d worldsheet CFTs. Indeed, there are no D-branes in heterotic string theory because one cannot introduce suitable boundary conditions on its chiral worldsheet theory\footnote{Note that although the type IIA string worldsheet is chiral in 2d, due to the opposite GSO projections, it is possible to introduce boundary conditions breaking part of the global symmetry (i.e. 10d Poincar\'e invariance).}. However, there is a remarkable construction of open heterotic strings in \cite{Polchinski:2005bg} in 10d flat space $SO(32)$ heterotic\footnote{In the $E_8\times E_8$ theory, the analogous constructions is possible, but it requires the presence of certain singularities in the geometry \cite{Polchinski:2005bg}, hence we skip it.} theory (see \cite{Alvarez-Garcia:2024vnr} for a recent discussion), as we now review. 

The point is that a heterotic $SO(32)$ string worldsheet can end on configurations of the $SO(32)$ gauge fields with non-trivial value for $\tr F^4$ on the $\IS^8$ surrounding the worldsheet boundary (i.e. the $\IS^8$ around the origin in the $\IR^9$ transverse to the string endpoint worldline). This can be shown to be consistent with flux conservation by checking the invariance of the action under gauge transformations of the 10d 2-form $B_2\to B_2+d\Lambda_1$. Indeed, the action contains the terms
\beqa
S_{B_2}=\int_{\Sigma_2} B_2 + \int_{10d} B_2\tr F^4,
\label{b2-action}
\eeqa
where the first term is the coupling of the string worldsheet $\Sigma_2$ and the second is the 10d 1-loop term required in the Green-Schwarz mechanism. Under gauge transformations, 
\beqa
\delta_{\Lambda_1} S_{B_2}=\int_{\partial\Sigma_2}\Lambda_1-\int_{10d} \Lambda_1 d\tr F^4=\int_{\partial\Sigma_2}\Lambda_1-\int_{10d} \Lambda_1 \delta_9(\partial \Sigma_2)=0\, .
\eeqa
Here, in the first equality we have used integration by parts, and in the next-to-last equality we have used $d\tr F^4=\delta_9(\partial \Sigma_2)$, where $\delta_9(\partial \Sigma_2)$ is a bump 9-form supported at the boundary $\partial\Sigma_2$ of the worldsheet (namely, by Gauss' law, $\tr F^4$ integrates to 1 over the $\IS^8$ around $\partial\Sigma_2$).

A second important ingredient in the discussion in \cite{Polchinski:2005bg} is that the gauge configuration carries away the excess of left- over right-moving fermions on the heterotic worldsheet. At the boundary of the open heterotic string, the left-moving fermions transition into fermions of the bulk theory which are carried in the radial direction away from the worldsheet boundary.

\subsubsection{The Cone construction}
\label{sec:cone-open-het}

The above configuration represents a non-trivial boundary for a 2d chiral theory, albeit in a theory embedded in a higher-dimensional theory. However, there is a simple way in which we can turn the system into a 2d configuration, which amounts to regarding a local flat space as a cone. This has been exploited in the context of building Local Dynamical Cobordism solutions in \cite{Angius:2022aeq}\footnote{Cone constructions of this kind have also been played a prominent role in the construction of SymTFTs (see \cite{Apruzzi:2021nmk}, also \cite{Schafer-Nameki:2023jdn} for a review), as well as in holography, starting from \cite{Klebanov:1998hh,Morrison:1998cs}.} and in fact it will produce dynamical cobordisms in our setup as well, c.f. section \ref{sec:dyn-cob-d3s}. We advance that, although the Cone Construction does not yield a boundary configuration for a genuine chiral 2d theory in this particular example of open heterotic strings, this construction will do the job in other examples in coming sections.

We hence regard the flat space local geometry around the open heterotic string worldsheet boundary in the previous section as a cone over $\IS^8$ (times the time direction along the boundary of the string worldsheet), and consider it from the perspective of the effective 2d theory obtained after a compactification on $\IS^8$. The cone configuration, in which the $\IS^8$ varies in the radial direction and shrinks at the origin, can thus be regarded as a dynamical cobordism solution of the 2d theory obtained after compactification on $\IS^8$, in analogy with \cite{Angius:2022aeq,Blumenhagen:2022bvh}, thus defining an ETW configuration ending spacetime. 

As in \cite{Angius:2022aeq}, the above configurations should be regarded merely as local descriptions near the ETW boundary, which can be part of a more involved global configuration, in which in particular the $\IS^8$ may have a finite size further away from the ETW boundary. A template for this behaviour is Witten's bubble of nothing \cite{Witten:1981gj}, in which the compactification $\IS^1$ has a constant asymptotic radius for, but locally near the bubble of nothing it is a polar angle which combines with the radial coordinate to parametrize a local $\IR^2$. We thus conceive our cone constructions in a similar spirit.

Let us thus consider the compactification of the 10d theory on $\IS^8$ (similar considerations can be made for more general spaces $\IX_8$). Since we want to match the cone construction of the previous section, we need to turn on a non-trivial $\tr F^4$ background on it. Note that from the 10d 1-loop coupling (\ref{b2-action}), the resulting 2d theory has a non-trivial tadpole for $B_2$ (the heterotic analogue of the tadpole in \cite{Sethi:1996es}), which has to be explicitly cancelled by the introduction of a fundamental string worldsheet, namely the first term in (\ref{b2-action}). This is just a rederivation of the flux conservation argument at the beginning of this section.

\begin{figure}[htb]
\begin{center}
\includegraphics[scale=.4]{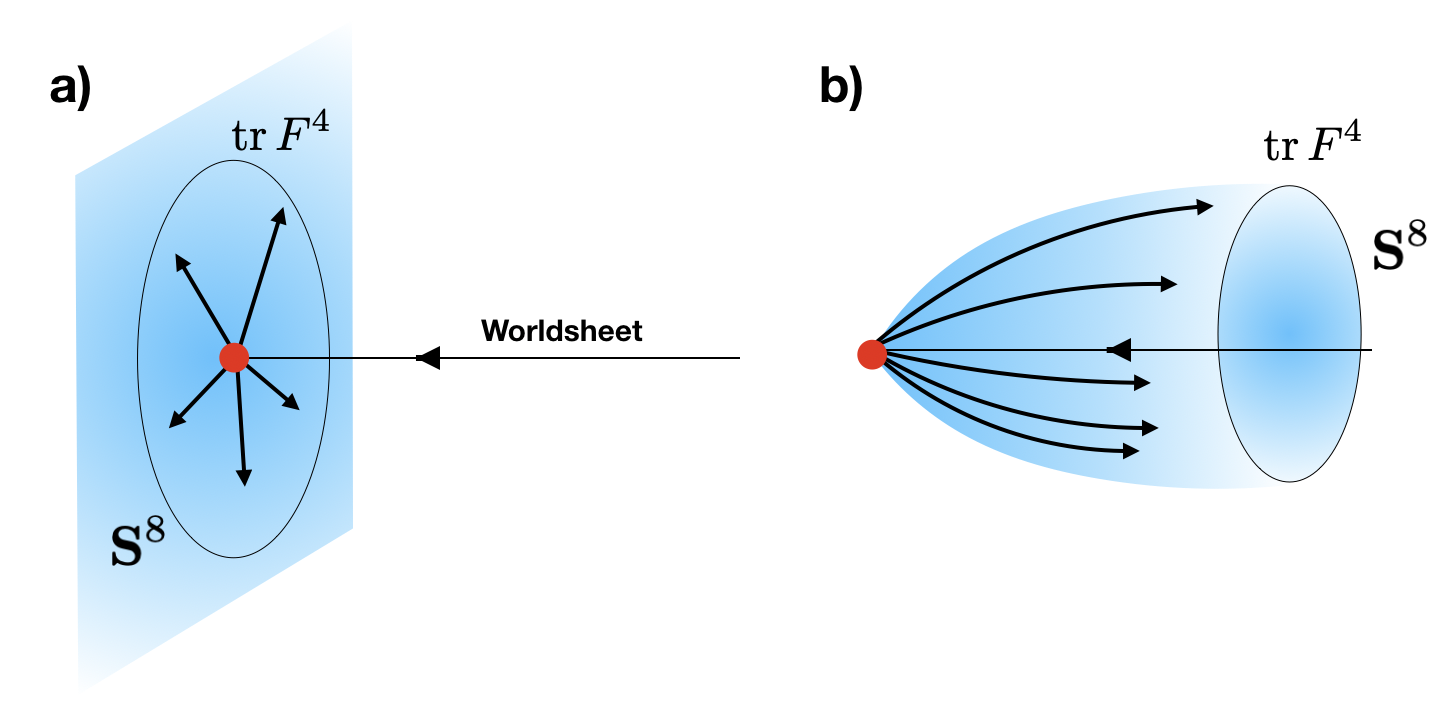}
\caption{\small a) Open heterotic string in flat space, with chiral fermions (denoted with arrows) ending on the boundary outflow as bulk zero modes as. b) The configuration in the Cone Construction: the theory upon compactification on $\IS^8$ describes a boundary configuration for a running 2d solution. However, the content of chiral fermions is non-chiral.}
\label{fig:outflow}
\end{center}
\end{figure}

Hence the worldsheet fields on this string worldsheet are now degrees of freedom of our 2d spacetime theory. Because they are chiral, one may, as mentioned above, have the expectation that we have a 2d chiral theory, which ends on a codimension 1 boundary where the $\IS^8$ shrinks. If true, this would actually be very striking, because the 2d theory on the worldsheet is anomalous and does not make sense by itself in the quantum theory. However we know that there are actually extra ingredients which come to the rescue, in the form of the fermion zero modes of the 10d gauginos in the presence of the gauge background. Indeed, the 10d chiral fermions in the adjoint of $SO(32)$ lead, upon compactification on $\IS^8$ with a non-trivial $\tr F^4$, to non-trivial 2d chiral fermions due to the index of the Dirac operator. The computation is essentially a reinterpretation of that in \cite{Polchinski:2005bg}, with the result that the chiral fermions coming from these zero modes cancel the chirality of the 2d fermions from the worldsheet, in fact in a trivial vector-like way. We thus end up with a non-trivial boundary configuration described as a dynamical cobordism ending spacetime, but for a 2d theory with a {\em vector-like} set of fermions. The situation is depicted in Figure \ref{fig:outflow}.

%%%%%%%%%%%
%%%%%%%%%%%

Despite the apparent failure to obtain a Chiral Cone Construction in this concrete example, we will continue exploiting the general strategy in coming examples. Namely, we consider branes or defects supporting chiral theories, introduce boundaries for them ensuring flux conservation, and regard the geometry around the boundary as a cone, and the configuration as a dynamical cobordism solution for the theory compactified on the base of the cone. Following these setups, we will eventually obtain boundary configurations in several large classes of models, discussed in later sections.

Let us finally mention that  there is interestingly a very explicit quantitative description of the above cone construction solution, in terms of a precise worldsheet theory for a non-critical heterotic string. In fact \cite{Kaidi:2023tqo} recently identified the boundary of the open heterotic string  as a non-supersymmetric 0-brane of heterotic theory, and provided the worldsheet description of the near horizon geometry, in terms of a (gapped) 2d $\NN=(1,1)$ sigma model describing an $\IS^8$ compactification with a non-trivial $\int_{\IS^8}\tr F^4$ background, and a radial direction with a linear dilaton background. As shown in \cite{Angius:2022mgh} such linear dilaton backgrounds turn into a dynamical cobordism in the Einstein frame. Hence our picture has a nice agreement with the setup in \cite{Kaidi:2023tqo}.

\subsection{Boundaries for Chiral 6d Theories from Open Heterotic NS5-branes}
\label{sec:sixdim}

The example in the previous section (see section \ref{sec:open-intersecting} and appendix \ref{app:more-intersecting} for other examples) illustrates an important point. In a defect supporting a chiral theory with an open worldvolume manifold, the boundary defines a transition in which the worldvolume ends and its localized degrees of freedom outflow as bulk modes. In this context, the cone construction allows to turn the system into a boundary configuration for the theory obtained as dimensional reduction on the base of the cone. However, this theory is non-chiral if the worldvolume degrees of freedom and the bulk degrees of freedom after the transition are of the same kind; in the previous examples, they both corresponded to chiral fermions transforming in the same representation of the gauge group, so they end up forming non-chiral pairs. Hence, the strategy to achieve a boundary for a genuinely chiral theory is to consider as starting point a chirality changing phase transition given by a process in which some brane ends on a boundary and the bulk modes outflowing from the boundary are of a totally different kind from the original worldvolume modes.

Chirality changing phase transitions have been a subject of active research in string theory and there is a good number of examples in the literature both in 6d \cite{Seiberg:1996vs,Aldazabal:1996du,Aspinwall:1996vc,Aspinwall:1997ye,Intriligator:1997kq,Blum:1997mm,Blum:1997fw,Brunner:1997gf,Hanany:1997gh} as well as in 4d \cite{Kachru:1997rs,Aldazabal:1997wi,Cvetic:2001nr}. In the following we illustrate the above picture with the paradigmatic case of the small instanton phase transitions in the $E_8\times E_8$ heterotic theory \cite{Seiberg:1996vs}. 

\subsubsection{The $E_8\times E_8$ heterotic NS5-brane chirality changing phase transition}
\label{sec:M5}

The $E_8\times E_8$ heterotic NS5-brane chirality changing phase transition (which is often discussed in terms of the Ho\v{r}ava-Witten M-theory uplift) is as follows \cite{Seiberg:1996vs}. Consider the 10d heterotic theory in flat spacetime, in the presence of one NS5-brane along the directions 012345. The worldvolume theory has 6d $\NN=(1,0)$ supersymmetry and contains a tensor multiplet, whose single real scalar parametrizes a Coulomb branch (corresponding to the position of the M5-brane in the Ho\v{r}ava-Witten interval $\IS^1/\IZ_2$ in the M-theory uplift), and one hypermultiplet, whose four real scalars parametrize a Higgs branch, the position of the NS5-brane (equivalently the M5-brane in the 11d lift) in the transverse dimensions 6789. By changing the vev of the scalar in the tensor multiplet one can reach the origin in the Coulomb branch (which corresponds to the 11d M5-brane reaching one of the Ho\v{r}ava-Witten boundaries) at which new massless degrees of freedom become light (M2-branes stretched between the M5 and the Ho\v{r}ava-Witten boundary) and the theory becomes strongly interacting. At this point the NS5-brane can be equivalently regarded as a zero-size small instanton (see \cite{Witten:1995gx} for the similar process for the $SO(32)$ heterotic), so it is possible to move into a Higgs branch, turning it into a finite size $E_8$ gauge instanton. The resulting spectrum of zero modes on the instanton can be obtained using the index theorem, which provide a spectrum 6d $\NN=(1,0)$ hypermultiplets. Consider the transition of $k$ 5-branes into an instanton background into an $SU(2)\subset E_8$, for simplicity. From the group theory decomposition
\beqa
E_8 \; & \quad \rightarrow\quad & E_7\times SU(2)\nonumber\\
{\bf 248} & \rightarrow & ({\bf 133},{\bf 1}) + ({\bf 56},{\bf 2}) + ({\bf 1}, {\bf 3})\, ,
\eeqa
the number of instanton fermion zero modes in the different $E_7$ representations given by the index theorem are
\beqa
\#{\bf 56}=(k-4)/2\quad ,\quad \#{\bf 1}=2k-3\, .
\eeqa
Hence, each single unit of instanton charge contributes a 6d $\NN=(1,0)$ half-hypermultiplet in the ${\bf 56}$ of $E_7$ and 2 singlets. 

Overall, the transition from the heterotic NS5-brane to the finite size gauge instanton has turned a spectrum with 1 tensor multiplet and 1 hypermultiplet into a total of 30 hypermultiplets. These 6d $\NN=(1,0)$ spectra are chiral, hence the transition is a chirality changing phase transition, albeit (and very remarkably) in a way compatible with the (highly restrictive) 6d anomaly cancellation conditions\footnote{Although we are focusing on configurations with non-compact 10 dimensions, it makes sense to consider the 6d anomalies, which in this context are considered as localized anomalies on the volume of the defects. They will become genuine 6d anomalies upon compactification e.g. on K3, or as in the cone construction in the next section.}. In particular, focusing on purely gravitational anomalies, the contribution from an overall number $V$ of vector multiplets, $H$ hypermultiplets and $T$ tensor multiplest is $H-V-29T$. This leads to the celebrated fact that a tensor multiplet can be traded for 29 hypermultiplets, as realized in the above phase transition. Let us also mention that, in addition, there are several other gauge anomalies that match in this transition, see \cite{Seiberg:1996vs} for details.

Note that for the $SO(32)$ heterotic there is a similar small instanton transition, but if the NS5-brane is at a smooth point in the transverse space, it does not involve tensor multiplets, and does not lead to chirality change. Hence, in the cone construction in the next section will lead to boundaries for non-chiral theories. The situation changes for 5-branes at singularities, as we explore in section \ref{sec:cone-singus}.

\subsubsection{The Open Heterotic NS5-brane and the Chiral Cone construction}
\label{sec:open-ns5}

Let us now exploit the above information to build an open NS5-brane configuration, in analogy with the open heterotic string in section \ref{sec:twodim}. 

The key point is that in the presence of NS5-branes with worldvolume $\Sigma_6$, the couplings for the 6-form $B_6$ dual of the 2-form field are
\beqa
\int_{\Sigma_6} B_6+\int_{10d} B_6\, (\tr F^2-\tr R^2)\, .
\eeqa
Equivalently, the modified Bianchi identity for the 3-form field strength is
\beqa
dH_3=\tr F^2 - \tr R^2+\delta_4(\Sigma_6)\, ,
\label{bianchi-ns5}
\eeqa
where $\delta_4(\Sigma_6)$ is a bump 4-form Poincar\'e dual to $\Sigma_6$. The above means that a 5-brane can have a 5d boundary if the latter acts as a source as $d(\tr F^2-\tr R^2)=\delta_5(\partial\Sigma_6)$, with $\delta_5$ a bump form for Poincar\'e dual to the boundary $\partial\Sigma_6$. In other words, if we denote by $\IX_4$ the geometry around the 5d boundary $\partial \Sigma_6$, we need
\beqa
\int_{\IX_4}  \tr F^2-\tr R^2=1\,.
\eeqa
There are several possibilities for this. The most direct is that the space transverse to $\partial \Sigma_6$ is smooth, hence locally $\IR^5$, and then $\IX_4=\IS^4$ at the topological level. Since $\int_{\IS^4}\tr R^2=0$, then we need a non-trivial gauge instanton bundle
\beqa
\int_{\IS^4} \tr F^2=1\, .
\eeqa
Another possibility is that the 5-brane boundary is located at the tip of a singular 5d transverse space, so that we can have some $\IX_4$ with non-trivial  second Pontryagin class. This will be explored in section \ref{sec:5branes-at-singus}, and here we consider just the case of $\IX_4=\IS^4$ with a non-trivial gauge bundle. 

It is easy to  construct a gauge background with instanton number 1 on $\IS^4$. One just picks an $SU(2)$ subgroup of the gauge group $E_8\times E_8$ or $SO(32)$, and regards $SU(2)$ as $\IS^3$, and builds the instanton background using the Hopf fibration of $\IS^7$ over $\IS^4$ with fiber $\IS^3$. We will not need this explicit construction and simply proceed  at an essentially topological level.

It is easy to describe what is happening at the boundary of the NS5-brane. When the chiral content of the 6d NS5-brane theory reaches the boundary, it encounters a non-trivial gauge background, which produces a set of bulk fermions (from the 10d gaugino zero modes) outgoing radially as 30 hypermultiplets\footnote{Actually, the configuration is in general not supersymmetric, but the counting of fermions is topological, so it works similarly and we abuse language and still use the susy jargon.\label{foot:non-susy}} (as $1/2\cdot{\bf 56}+ 2\cdot{\bf 1}$ of $E_7$) and carrying away the anomaly. The total charge under $B_6$ (i.e. the $H_3$ flux) is conserved, and so is anomaly, albeit in a very non-trivial way, because of the trading of 1 tensor for 29 hypers. This is a key different with respect to the open heterotic string in section \ref{sec:twodim}, and impacts crucially in the cone construction, to which we now turn.

Let us now regard the $\IR^5$ transverse to $\partial \Sigma_6$, as a cone over $\IS^4$. We can regard this as a dynamical cobordism of the 6d theory obtained upon compactification of the 10d theory on $\IS^4$. On this $\IS^4$, the NS5-brane is sitting at a point, and leads to a 6d $\NN=(1,0)$ tensor and a hyper. The 5-brane charge is cancelled by a gauge background
\beqa
\int_{\IS^4}\tr F^2=-1\, .
\eeqa
The change of sign is due to a change in the orientation of the $\IS^4$ when regarded in flat space or in the cone. It implies we get fermions of chirality opposite to that of hypers (reflecting the compactification is non-susy), Namely, we get  `opposite chirality' hypers in the $1/2\cdot{\bf 56}+ 2\cdot{\bf 1}$. 

Overall, the total content is (very!) chiral, but non-anomalous, with the anomaly from the tensor cancelling against the 29 `opposite chirality' hypers. The configuration describes a dynamical cobordism in a 6d chiral non-anomalous 6d theory, in which the scalar parametrizing the $\IS^4$ size runs until it shrinks to zero size, ending spacetime. We moreover have a fairly good microscopic understanding of the ETW configuration, in terms of the NS5-brane boundary. The theory admits a boundary, with effective boundary conditions relating wildly different fermion fields, thanks to a non-trivial mechanism gapping the chiral non-anomalous content, necessarily at strong coupling. 

One interesting perspective of the cone construction is that, in the same way its use to construct SymTFTs allows an efficient way to study singular configurations by means of a smooth compactification, in our present setup it may serve to get further information about the strongly coupled regime of the transition between the NS5-brane and the instanton. We will say a bit more on this in section \ref{sec:symtft}.

\subsubsection{A related Open M5-brane Chiral Cone Construction}
\label{sec:open-m5}

The above cone construction describing open heterotic NS5-branes is closely related to the system studied in \cite{Bergshoeff:2006bs} in supergravity, and recently revisited in \cite{Kaidi:2023tqo} from the viewpoint of a worldsheet description. In this section we show that this system, when regarded as a cone construction, also corresponds to a boundary configuration for a chiral compactification of the $E_8\times E_8$ heterotic theory.

Let us revisit the setup in \cite{Bergshoeff:2006bs,Kaidi:2023tqo}. Consider an M5-brane extending in the $\IS^1/\IZ_2$ interval between the two boundaries of the Ho\v{r}ava-Witten theory, and turning into an instanton of the $E_8$ at each of the boundaries. Equivalently, an $E_8$ instanton on a first boundary turns into an M5-brane, which travels along the interval, and turns into an $E_8$ instanton of the second boundary. The system is morally a double copy of the chirality changing phase transition in the previous sections. In the heterotic string limit of small M-theory interval size, we just have one $E_8$ instanton turning into an instanton of the second $E_8$. This configuration was discussed in the supergravity approximation \cite{Bergshoeff:2006bs}, while \cite{Kaidi:2023tqo} identified the transition region as a non-supersymmetric heterotic 4-brane and provided an explicit worldsheet description of its near horizon regime.

We would like to pursue the latter 4-brane perspective with emphasis in regarding it as a cone construction. The geometry around the 4-brane can be regarded as a cone over $\IS^4$, on which there is a non-trivial instanton background with instanton number $(1,-1)$ embedded in $SU(2)\times SU(2)\subset E_8\times E_8$. The near horizon regime was shown in \cite{Kaidi:2023tqo} to be given by 2d (gapped) $\IS^4$ sigma model, times a quotient of a 4 Majorana-Weyl fermion $SO(4)$ free theory and an $E_7\times E_7$ current algebra CFT (describing the unbroken gauge symmetry), times a linear dilaton theory describing the radial coordinate.

From the cone construction perspective, the system is providing a dynamical cobordism for the 6d theory obtained by compactifying the 10d heterotic theory on $\IS^4$ with a non-trivial instanton background in $SU(2)\times SU(2)\subset E_8\times E_8$. Namely, a boundary configuration for the 6d chiral theory with gauge symmetry $E_7\times E_7$ and chiral matter content given by a set of fermions charged under the first $E_7$, with multiplicities dictated by the index theorem, and the corresponding opposite chirality fermions charged under the second $E_7$. Hence, this simple cone construction provides a boundary configuration for a chiral 6d theory. 

Let us remark that, even though our derivation involved a double use of the chirality changing phase transition of the previous sections, in the final chiral cone construction that complicated physics is all hidden at the tip of the cone. In fact, it is possible to propose a simpler description of the boundary conditions at the tip of the cone in terms of exchange of left- and right chiralities, with a simultaneous exchange of the two $E_8$'s, which is a gauge symmetry of the 10d theory. We will encounter similar examples in the 4d context in section \ref{sec:cone-singus}.

We finally note that, although the resulting full configuration is non-supersymmetric and would seem complicated, its behaviour near the tip is explicitly described by the near horizon worldsheet theory in \cite{Kaidi:2023tqo}. Moreover, its description of the radial direction as a linear dilaton theory, implies as in \cite{Angius:2022mgh} that in the Einstein frame it corresponds to a dynamical cobordism in which the dilaton runs and blows up at a finite spacetime distance. This nicely reproduces our intuition that the cone construction correspond to dynamical cobordisms of the
theory after compactification on the base of the cone. The relation with dynamical cobordisms will be explicitly recovered, in an analogous class of cone constructions, in section \ref{sec:dyn-cob-d3s}.

\subsection{Chirality Changing Phase Transitions from 5-branes at Singularities}
\label{sec:5branes-at-singus}

In this section we briefly point out that the 5-brane chirality changing phase transition in section \ref{sec:M5} has several generalizations, obtained by locating the 5-brane at the tip of an orbifold singularity. This has been efficiently studied for D5-branes at $\IC^2/\IZ_N$ singularities in \cite{Intriligator:1997kq,Blum:1997mm,Blum:1997fw}
(see also similar results and generalization from Hanany-Witten brane constructions \cite{Brunner:1997gf,Hanany:1997gh} and from the perspective of F-theory on CY3 in \cite{Aspinwall:1996vc, Aspinwall:1997ye} and in the recent \cite{DelZotto:2022ohj,DelZotto:2022xrh}).

For concreteness, we will focus on a particular illustrative example, based on the chirality changing phase transition for type I D5-branes at the $\IC^2/\IZ_2$ singularity, studied in \cite{Intriligator:1997kq} (see \cite{Aspinwall:1996vc} for an earlier derivation in F-theory on CY3, and \cite{Brunner:1997gf,Hanany:1997gh} for a derivation in a T-dual type I' theory with D6-branes suspended among NS5-branes). The discussion generalizes straightforwardly to more general cases, which we leave as an exercise for the interested reader.

Consider type I theory on $M_6\times \IC^2/\IZ_N$, with the generator $\theta$ of $\IZ_N$ acting as $\theta:(z_1,z_2)\to (e^{2\pi i/N}z_1,e^{-2\pi i/N}z_2)$, which preserves 8 supersymmetries, i.e. 6d $\NN=1$ at the tip of the singularity. As explained in \cite{Polchinski:1996ry} there are two choices of the orientifold action on the orbifold twisted sector: the choice without vector structure, which gives a 6d $\NN=1$ hypermultiplet in the twisted sector, and breaks the D9-brane symmetry down to $U(16)$, and the choice with vector structure, which produces a  tensor multiplet, and breaks the D9-brane symmetry down to $SO(w_0)\times SO(w_1)$,  where these integers satisfy $w_0+w_1=32$, and define asymptotic holonomy of the D9-brane gauge bundle. We focus on the latter case, i.e. with vector structure, and for simplicity we choose $w_0=32$, $w_1=0$, so the unbroken symmetry is $SO(32)$.

We can now locate a number of D5-branes at the tip of the singularity, without further breaking of supersymmetry, so we get a 6d $\NN=1$ gauge theory on their worldvolume. The spectrum is
\beqa 
&USp(2k)\times USp(2k-8) &\nonumber\\
&(\fund,\fund) +16(\fund, {\bf 1})+1\cdot {\bf T} \, ,&
\label{spec-z2}
\eeqa
where ${\bf T}$ is the tensor multiplet, and the 16 fundamentals arise from the D5-D9 open string sector and actually correspond to one half-hypermultiplet in the $(\fund;{\bf 32})$ of $USp(2k-8)\times SO(32)$, with the latter regarded as a global symmetry from the 6d perspective. 

In the limit of strong coupling of the $USp(2k-8)$ theory, which is the origin of the Coulomb branch for the tensor multiplet, there exists a chirality changing phase transition, in which this gauge factor disappears and so do the bifundamental hypermultiplet and the tensor multiplet, while there appears hypermultiplets in the $\asymm+2\cdot{\bf 1}$ of the $USp(2k)$ factor, which parametrize a Higgs branch. The theory is thus
\beqa 
&USp(2k) &\nonumber\\
&\asymm +16\fund+ 2\cdot {\bf 1}\, . &
\label{spec-z2-prime}
\eeqa
The anomalies of the theories before and after the transition fully agree, in particular again effectively trade 1 tensor multiplet for 29 hypermultiplets. For instance, for $k=4$, the initial theory is simply $USp(8)$ with 16 hypers in the fundamental and a tensor multiplet, and the whole transition amounts to removing the tensor multiplet and replacing it by a hypermultiplet in the ${\bf 27}$ of $USp(8)$ plus two singlets.

We can now move into the Higgs branch in particular giving vevs to the 16 fundamentals, which corresponds to dissolving the D5-branes as gauge instantons of the D9-brane theory. This breaks the $USp(2k)$ group completely, and the $SO(32)$ down to some subgroups depending on the gauge embedding of the instantons. Embedding them as an instanton number $k$ background in an $SU(2)\subset SO(32)$ for simplicity, we can use the decomposition
\beqa 
SO(32)\; & \;\to\; & \; SU(2)\times SU(2)\times SO(28)\nonumber\\
{\bf 496} &\to &  ({\bf 3},{\bf 1},{\bf 1})\,+\, ({\bf 1},{\bf 3},{\bf 1})\,+\,({\bf 1},{\bf 1},{\bf 378})\,+\, ({\bf 2},{\bf 2},{\bf 28})\, ,
\eeqa 
and get the  hypermultiplet spectrum from the index theorem, which gives
\beqa
&\#_{({\bf 2},{\bf 28})}=(k-4)/2\quad ,\quad \#_{({\bf 1},{\bf 1})}=2k-3\, .&
\eeqa 
The transition is again compatible with the structure of anomalies, once the change in vector multiplets from the breaking $SO(32)\to SO(28)\times SU(2)$ is taken into account.

For completeness, let us describe this transition from the perspective of Hanany-Witten brane configurations in type I' theory, which is obtained upon T-dualizing the system above along the $\IS^1$ corresponding to the $U(1)$ orbit $(z_1,z_2)\to (e^{i\varphi}z_1,e^{-i\varphi}z_2)$ in $\IC^2/\IZ_N$ (see \cite{Brunner:1997gf,Hanany:1997gh}, also \cite{Park:1998zh} for a 4d $\NN=2$ version). We have type I' theory, i.e. IIA on $\IS^1/\IZ_2$, with a $\IZ_2$ orientifold quotient introducing O8$^-$ planes at the two fixed loci. The $\IC^2/\IZ_2$ orbifold is mapped to two NS5-branes in the covering $\IS^1$, and the choice with vector structure corresponds to having them at orientifold image points away from the O8$^-$-planes. The choices of $w_0$, $w_1$ describe the distribution of the 32 D8-branes in the two intervals separated by the NS5-branes, i.e. on top of each of the O8$^-$-planes, so the choice $w_0=32$, $w_1=0$, leads to the 32 D8-branes on top of one O8$^-$-plane, leaving the other empty. We now stretch $2k$ D6-branes suspended between the NS5-branes in the interval passing through the occupied O8$^-$-plane, and $2k-8$ D6-branes between the NS5-branes but on the interval passing through the empty O8$^-$-plane. The spectrum of the 6d $\NN=1$ theory is (\ref{spec-z2}), with the tensor multiplet corresponding to the position of the NS5-branes on the $\IS^1$. We can now move the NS5-brane and its image on top of the empty O8$^-$-plane, by tuning the scalar in the tensor multiplet. Then there exists a phase transition, corresponding to moving the NS5-branes, as two independent objects, along the O8$^-$-plane and off the D6-branes. The set of left-over D6-branes leads to the $USp(2k)$ theory with the 2-index antisymmetric hypermultiplet, while the tensor multiplet has disappeared because the NS5-brane position in $\IS^1$ is fixed. The positions of the NS5-branes away from the D6-branes parametrize 2 hypermultiplet singlets, and the rest of the Higgs branch is parametrized by the antisymmetric matter and the bifundamentals, whose effect was discussed in the previous paragraph. This picture of the transition in terms of brane motions is completely general and applies to the infinite classes of 6d $\NN=1$ theories from type I D5-branes at $\IC^2/\IZ_N$ singularities with and without vector structures.

Let us now go back to our particular $\IC^2/\IZ_2$ example and carry out a Chiral Cone construction based on the above chirality changing phase transition. Consider type I theory on $\IC^2/\IZ_2$ and take the 5d space $\IR\times\IC^2/\IZ_2$, with $\IR$ parametrized by one of the 6d Poincar\'e invariant coordinates, say $x^5$, and regard it as a cone over a 4d base $\IS^4/\IZ_2$. The $\IZ_2$ action has two fixed points on $\IS^4$, locally of the form $\IC^2/\IZ_2$. We now turn on an instanton number $-k$ gauge background in $SU(2)\subset SO(32)$, and locate $k$ D5-branes at one of the $\IC^2/\IZ_2$ singularities, so as to be compatible with untwisted RR tadpole cancellation for this compactification. The flip of the gauge bundle instanton background is due to the orientation flip between the coordinate $x^5$ and the radial coordinate of the cone for the two singularities. 

The spectrum is given by 
\beqa
{\rm Vectors:}\quad &\; USp(2k)\times USp(2k-8)\times SU(2)\times SO(28)& \nonumber\\
{\rm Hypers:}\quad & \;(\fund,\fund;{\bf 1},{\bf 1}) +(\fund, {\bf 1};{\bf 2}, {\bf 1})+ \frac 12(\fund, {\bf 1};{\bf 1}, {\bf 28})& \nonumber\\
{\rm Hypers':} &\frac{(k-4)}{2}\,({\bf 1},{\bf 1};{\bf 2},{\bf 28})  + (2k-3) \,({\bf 1},{\bf 1};{\bf 1},{\bf 1})&  \nonumber\\
{\rm Tensors:}\quad &\;1\cdot {\bf T} \, .& 
\label{z2-instanton-transition-theory}
\eeqa
The second line is the group theory decomposition of the hypermultiplet content in (\ref{spec-z2}), while the hypers' in the second indicate `opposite chirality' hypermultiplets. We recall that our use of susy jargon is merely for convenience, c.f. footnote \ref{foot:non-susy}.

In analogy with section \ref{sec:open-ns5}, the above compactification admits a running dynamical cobordism solution microscopically given by the flat space solution regarded as a cone. The physics at the origin is the chirality changing phase transition described above, namely the transformation of the dynamical tensor multiplet of a pointlike instanton at $\IC^2/\IZ_2$ into a set of hypermultiplets associated to their fattening into a gauge instanton. Hence the dynamical cobordism provides a boundary configuration for the 6d chiral theory (\ref{z2-instanton-transition-theory}).

We note that, even though the 6d theory has a highly non-supersymmetric, the final running solution describing the dynamical cobordism is supersymmetric, as it secretly corresponds to the system of D5-branes at an orbifold of flat space. The fact that dynamical cobordism solutions may enjoy more supersymmetry than the effective theory is familiar from several other examples, see e.g. \cite{Buratti:2021yia}.

We again emphasize that this construction technique generalizes straightforwardly to other chirality changing phase transitions of 6d $\NN=1$ theories, and leave further examples for the interested reader.

\subsection{Relation to SymTFTs}
\label{sec:symtft}

In this section we would like to highlight an interesting connection. We have exploited the cone construction to regard interesting phenomena occurring in a region localized in a $(d-1)$-dimensional subspace of spacetime in terms of the evolution in the $d$-dimensional theory obtained by compactification on the angular manifold around it, namely on the base of the cone describing its transverse space. This technique has been applied, at the topological level, in a different context related to generalized symmetries in quantum field theory and string theory (see \cite{McGreevy:2022oyu,Brennan:2023mmt,Gomes:2023ahz,Shao:2023gho,Schafer-Nameki:2023jdn,Bhardwaj:2023kri,Iqbal:2024pee} for reviews), as follows. 

For a $(d-1)$-dimensional field theory (possibly coupled to gravity), the set of generalized symmetry generators and of generalized charged operators can be encoded as the set of topological operators in a $d$-dimensional gapped topological field theory, known as the SymTFT (or, more generally, if some degree of non-topological sectors is allowed, Symmetry Theory). The SymTFT is given by a $d$-dimensional sandwich with two boundaries separated by an interval, one describing the local degrees of freedom of the original $(d-1)$-dimensional theory (referred to as {\em relative} theory, in the sense of \cite{Freed:2012bs}), and a second one providing the gapped topological boundary conditions for the SymTFT fields. The actual (or {\em absolute}) theory, including the global topological information, is recovered by collapsing the SymTFT interval. 

For $(d-1)$-dimensional theories which can be constructed as localized sectors in string theory or M-theory, a useful tool to derive the corresponding $d$-dimensional SymTFT \cite{Apruzzi:2021nmk} is to regard the transverse space as a cone, and to perform the dimensional reduction of the topological sector of the 10d string theory or 11d M-theory over the base of the cone\footnote{The approach is clearly inspired in the similar role played by cones in holography \cite{Klebanov:1998hh,Morrison:1998cs}, as pioneered in the generalized symmetries of 4d $\NN=4$ $SU(N)$ theory using holography in \cite{Witten:1998wy}.}. The resulting $d$-dimensional topological field theory is the SymTFT, with the physical theory realized at the tip of the cone, and the topological boundary given by the asymptotic boundary conditions at infinity in the cone. Moreover, the different topological operators are realized as (the topological sector of) different branes of the compactification; specifically, generalized symmetry operators correspond to branes at infinity, parallel to the boundaries, while charged topological defects arise from branes stretching in the radial direction of the cone.

It is clear that our Cone Constructions are based on a similar viewpoint, and in particular they should be closely related if our Cone Construction is truncated to its topological sector. In this perspective, in our above examples  the $(d-1)$-dimensional  physical theory at the tip of the cone corresponds to the boundary of the relevant brane (such as the open heterotic string or the 5-branes), while the SymTFT is the topological sector of the 10d string theory compactified on the corresponding sphere, with the corresponding fluxes, branes and any other ingredients. 

Specifically, our construction shows that the SymTFT of the open heterotic string boundary in section \ref{sec:twodim} is the topological sector of the compactification of 10d heterotic string on $\IS^8$ with one explicit fundamental string at a point and $-1$ units of gauge `flux' $\int_{\IS^8}\tr F^4=-1$. This is actually related to the comment  in section \ref{sec:cone-open-het} about \cite{Kaidi:2023tqo}, where an explicit worldsheet description of this configuration around the 0-brane, in the near horizon limit was provided. It would be interesting to explore the topological structures of this cone construction and possibly uncover novel features about the boundary of the open heterotic string.

Similarly, for the open heterotic NS5-brane in flat space, in section \ref{sec:open-ns5}, the boundary of the NS5-brane is a 4-brane, whose SymTFT is the topological sector of the compactification of 10d heterotic string on $\IS^4$ with one explicit NS5-brane and $-1$ units of instanton charge $\int_{\IS^4}\tr F^2=-1$. In this case, the 4-brane solution presented in \cite{Kaidi:2023tqo} actually corresponds to a system where two such chirality changing phase transitions are combined, as emphasized in section \ref{sec:open-m5}, and the asymptotic cone contains no explicit NS5-branes, but a pair of opposite charge instantons under the two $E_8$ gauge factors (or rather, $SU(2)$ subgroups thereof). In any event, we expect that the topological structure of the chirality changing phase transition of these NS5-branes (and possibly those from singular geometries) can be unravelled using the SymTFT constructions we have described.

One general observation about the $d$-dimensional theories arising from compactification on the base, is that, when the system describes a boundary configuration for a genuine chiral theory, namely when we have a genuine Chiral Cone Construction, the $d$-dimensional theory is not trivially gappable. This is simply because the $d$-dimensional chiral theory is part of the massless spectrum of the theory after compactification, and being chiral but non-anomalous, cannot be trivially gapped.

Hence the use of the familiar term SymTFT, which assumes a gapped topological field theory, involves  a slight abuse of language. Indeed, we should rather speak about a Symmetry Theory, which contains some non-topological degrees of freedom, yet whose topological sector is relevant to the generalized symmetries and its operators. The need to generalized beyond the naive concept of SymTFT has occurred in various contexts, leading to nover setups such as SymTrees \cite{Baume:2023kkf}, Nested SymTFTs \cite{Cvetic:2024dzu} or SymTFT Fans \cite{GarciaEtxebarria:2024jfv}. In particular, the presence of branes stretching in the radial direction in the cone and carrying the non-topological degrees of freedom associated to a chiral sector, suggests an interesting connection with the flavour branes and their realization in Symmetry Theories in \cite{Cvetic:2024dzu}. Hence, it is an interesting question how to deal with the Symmetry Theory associated to these systems. We leave this interesting question for the future, and now turn to the study of 4d theories.

\section{Boundaries for 4d Chiral Theories}
\label{sec:fourdim}

In order to construct boundary configurations for 4d chiral theories, one may proceed by considering the 4d version of chirality changing phases transitions, which has been considered for heterotic compactifications on CY3 \cite{Kachru:1997rs} (see also \cite{Aldazabal:1997wi}). We will however focus on alternative approaches, realized in terms of D-branes. 

In this section we develop several strategies to use the cone construction over chiral D-brane models to build boundary configurations for 4d chiral theories. After an initial discussion of cone constructions over intersecting D6-branes, we focus on systems of D3-branes at singularities, and obtain large classes of working models in this last setup.

\subsection{Cones over D6-brane intersections}
\label{sec:open-intersecting}

In this section we study configurations of intersecting D6-branes, such that the 4d chiral fermions at their intersection are defined on a half-space, and carry out the cone construction around their 3d boundary. The specific example will eventually lead to a non-chiral theory upon this cone construction, albeit in a non-trivial and interesting way. It will thus serve as stepping stone in the construction of successful classes of models in coming sections.

There are two key ingredients in the construction of the 4d chiral fermion defined on a defect with boundary, which we study in turn. 

\subsubsection{4d chiral fermions from intersecting D6-branes}
\label{sec:inters}

In flat 10d type IIA theory a configuration of two stacks of $N_1$ and $N_2$ D6-branes intersecting over a 4d subspace of their worldvolumes, leads to a 4d chiral fermion transforming in the bifundamental\footnote{\label{foot:chirality} Recall that the chirality of the fermion (or equivalently, the fact of getting this bifundamental vs its conjugate) is determined by the relative orientation defined by the two intersecting 3-planes spanned by the D6-brane stacks (besides the Poincaré invariant 4d).} $(\fund_1,\antifund_2)$ \cite{Berkooz:1996km}. This is the setup which underlies model building via intersecting D6-brane worlds \cite{Blumenhagen:2000wh,Aldazabal:2000dg, Aldazabal:2000cn} (see \cite{Ibanez:2012zz} for review and references).

More explicitly, let the $N_1$ D6$_1$-branes span the directions 0123 and a 3-plane $\Pi_1$ in the remaining $\IR^6$, and let the $N_2$ D6$_2$-branes span 0123 and a 3-plane $\Pi_2$ in the remaining $\IR^6$. Even more explicitly, consider the $SO(6)$ rotation in $\IR^6$ that takes $\Pi_1$ to $\Pi_2$, and changes the basis of coordinates in spacetime so that the rotation is block diagonal. In this basis, the $\IR^6$ splits into $\IR^2\times\IR^2\times \IR^2$, and the 3-planes spanned by the D6-branes look like the product of three real lines in the three 2-planes. Let us denote $\theta_i$ the rotation angle that takes the line of the D6$_1$-branes to that of the D6$_2$-branes in the $i^{th}$ 2-plane. The configuration preserves 4 susys ($\NN=1$ in the 4d intersection) if the $SO(6)$ rotation is in $SU(3)$, in other words
\beqa
\theta_1\pm\theta_2\pm \theta_3=0\; {\rm mod} \; 2\pi \, .
\eeqa
The open string spectrum at the intersection is a 4d chiral fermion in the bifundamental representation $(\fund_1,\antifund_2)$ of the $U(N_1)\times U(N_2)$ on the D6-branes. In the susy $SU(3)$ case, there are also massless complex scalars that complete the spectrum to a 4d $\NN=1$ chiral multiplet.

In cases where the amount of supersymmetry is not important (e.g. topological aspects), we will use a simple example of 3-planes, and take e.g. the D6$_1$-branes to span the directions 0123456, and the D6$_2$-branes to span the directions 0123789. In this case, in the $\IR^2$'s parametrized by 47, 58, 69, respectively, the D6-branes are at angles $\theta_i=\pi/2$, which does not preserve susy. But the key topological ingredients, e.g. the presence of the localized 4d chiral fermion in the $(\fund_1,\antifund_2)$ are still present.

Notice that the localized anomaly of the above 4d fermion in the $(\fund_1,\antifund_2)$ is cancelled by an anomaly inflow mechanism \cite{Green:1996dd}. The consistency of inflows is the analogue in this setup of the conservation of fluxes for open branes in previous sections.

\subsubsection{Open D6-branes}
\label{sec:opend6}

In order to define boundaries for the above defect supporting the 4d bifundamental fermion, we intend to put boundaries in the above configurations of intersecting D6-branes. This first requires the discussion of how to define boundaries for a single isolated stack of D6-branes. In particular we explore D6-branes ending on NS5-branes (for D6-branes ending on D8-branes, see footnote \ref{foot-D8}).

As discussed in \cite{Hanany:1997sa}, in type IIA in the presence of a Romans mass $m$, an NS5-brane must emit $m$ semi-infinite D6-branes. Alternatively, in the presence of a Romans mass $m$, a set of $m$ D6-brane can end on one NS5-brane. A simple way to derive this, in analogy with the argument in sections \ref{sec:open-het}, \ref{sec:open-ns5}, is the following. We demand invariance of the action of the configuration under a gauge transformation of the RR 7-form $C_7\to C_7+d\Lambda_6$. The relevant pieces in the action are
\beqa
S_{C_7}=\int_{\Sigma_7} C_7+m\int_{10d}  C_7H_3\, .
\eeqa
The first term is the coupling of a D6-brane spanning a submanifold $\Sigma_7$, and the second is a topological coupling of Romans massive IIA theory. Its gauge variation is
\beqa
\delta_{\Lambda_6} S_{C_7}=\int_{\Sigma_7} d\Lambda_6+m\int_{10d}  d\Lambda_6 H_3=\int_{\partial \Sigma_7}\Lambda_6- m\int_{10d} \Lambda_6dH_3\, .
\eeqa
So, if the D6-branes end on an NS5-brane, we have $dH_3=\delta_4(\partial\Sigma_7)$, with $\delta_4(\partial\Sigma_7)$ a bump form Poincar\'e dual to $\partial\Sigma_7$, and hence
\beqa
\delta_{\Lambda_6} S_{C_7}=\int_{\partial\Sigma_7}\Lambda_6- m\int_{10d} \Lambda_6\delta_4(\partial\Sigma_7)=0\, .
\eeqa
An equivalent derivation is that the 10d coupling $m H_3 C_7$ turns the $U(1)$ gauge symmetry of $C_7$ into a discrete $\IZ_m$ symmetry, so that the electrically charged objects (D6-branes) are conserved only modulo $m$ \cite{Berasaluce-Gonzalez:2012awn}. The NS5-brane is the operator which must be dressed with electric D6-brane operators to be gauge invariant. Similarly, the emission effect can be regarded as a Freed-Witten anomaly on the NS5-brane \cite{Freed:1999vc,Maldacena:2001xj} (see also \cite{Berasaluce-Gonzalez:2012awn}), or equivalently from a D6-brane creation effect upon bringing $m$ D8-branes from infinity and crossing them over the NS5-branes as domain walls to introduce the Romans mass.

\subsubsection{4d chiral fermion on an intersection with boundary}
\label{sec:4dferm-bdry}

Consider an intersecting brane configuration, with a stack of $N_1$ D6$_1$-branes along 0123 456, and a second one of $N_2$ D6$_2$-branes along 012 789, the latter of semi-infinite extent in the direction 3 with the D6-branes ending on one NS5-brane located at $x^3=0$ and spanning 012 789. For this to be consistent we turn on a Romans mass $m=N_2$, as discussed in the previous section. The D6$_1$-branes are instead taken infinite (see appendix \ref{app:more-intersecting} for the case of both kinds of D6-branes being semi-infinite). 

\begin{figure}[htb]
\begin{center}
\includegraphics[scale=.4]{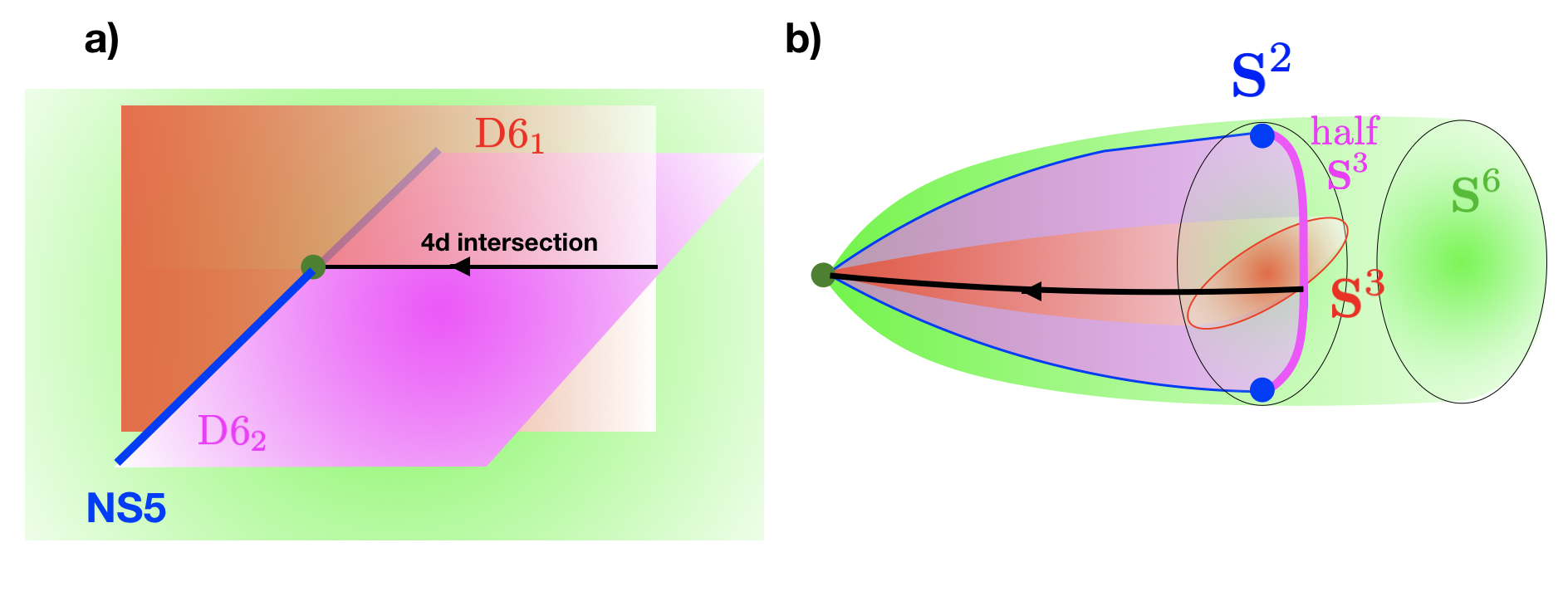}
\caption{\small Stack of infinite D6-branes intersecting a stack of semi-infinite D6-branes (ending on an NS5-brane) over 4d half-space in flat 10d space. a) The view in flat space. b) The cone perspective:  the green cone represents the flat 7d space spanned by 3456789, regarded as a cone over $\IS^6$. The red area is the  D6$_1$-brane worldvolume along the 4d space spanning 3456, regarded as a cone over $\IS^3_1$; the magenta area is the semi-infinite D6$_2$-brane, along 3789, which is a cone over a half $\IS^3_2$; it ends on the NS5-brane, in blue, which spans 789, regarded as a cone over $\IS^2$. The $\IS^3_1$ and the half $\IS^3_2$ intersect at one point on the $\IS^6$, so the intersection spans the black line along the radial direction. For clarity, the cone over $\IS^6$ has been extended slightly longer than the other cones.}
\label{fig:half}
\end{center}
\end{figure}

The spectrum gives 7d gauge fields on the D6$_1$-branes, 7d gauge fields on the half-space on the D6$_2$-branes, and a 4d chiral fermion on a half-space corresponding to the intersection, see Figure \ref{fig:half}a. 

%%%%%%%%%%%
%%%%%%%%%%%

Although it seems that we are harmlessly combining the two ingredients introduced in the previous section, it is clear that the above configuration cannot be complete, as can be argued in several ways. For instance, there is no consistent inflow mechanism, since the inflow from the D6$_1$-branes to the 4d intersection must suddenly stop when the intersection ceases to exist. Related to this, in the open heterotic string example we saw that chiral fermions reaching a boundary must outflow in some way, which is not obvious in the above description. Finally, if we turn the geometry into a cone, the missing fermions degrees of freedom imply we get an effective anomalous theory.

For illustration, let us be more explicit about this last argument, by performing the cone construction, depicted in Figure \ref{fig:half}b. We regard the $\IR^7$ spanned by 3456789 as a cone over $\IS^6$. The D6$_1$-branes span the directions 3456, so they span a cone over an $\IS^3\subset \IS^6$ defined by $(x^3)^2+(x^4)^2+(x^5)^2+(x^6)^2=R^2$. The NS5-brane spans the direction 789, namely a cone over an $\IS^2\subset \IS^6$ defined by $(x^7)^2+(x^8)^2+(x^9)^2=R^2$. The $\IS^2$ and $\IS^3$ do not intersect but are linked on $\IS^6$. The D6$_2$-branes span the direction 789 and are semi-infinite in 3 (because they end on the NS5-brane), so the span a cone over $(x^3)^2+(x^7)^2+(x^8)^2+(x^9)^2=R^2$ with $x^3>0$, namely a half-$\IS^3$ bounded by the $\IS^2$ wrapped by the NS5-brane. The half-$\IS^3$ of the D6$_2$-branes intersects the $\IS^3$ of the D6$_1$-branes at one point, $x^4=x^5=x^6=x^7=x^8=x^9=0$, $x^3=R$; the cone over this point is the direction supporting the 4d fermion over the semi-infinite radial direction.

So in the compactification of the 10d theory on $\IS^6$ we have D6$_1$-branes wrapped on an $\IS^3_1$ and D6$_2$-branes wrapped on a half-$\IS^3_2$ ending on an NS5-brane wrapped on the $\IS^2$ at the equator of $\IS^3_2$. The two sets of D6-branes intersect at one point in $\IS^6$ leading to one 4d chiral fermion in the $(\fund_1,\antifund_2)$. Hence, the resulting 4d theory is anomalous, making it manifest that we are missing some degrees of freedom.

\subsubsection{The missing D4-branes}
\label{sec:missing}

The appearance of anomalies suggests that the configuration in the previous section must be inconsistent as it stands. In fact, it is easy to see why, and to solve the problem. 

Consider the intersection of the NS5-brane and the D6$_1$-branes, namely the locus parametrized by 012 and located at the origin in 3456789. This locus is real codimension 4 in the D6$_1$-brane worldvolume. Then, in the D6$_1$-brane worldvolume, we can take an $\IS^3$  which surrounds the NS5-brane , namely the angular part of the $\IR^4$ spanning 3456. Since the NS5-brane is magnetically charged under the NSNS 2-form
\beqa
\int_{\IS^3}H_3=1\, .
\eeqa
So, if we excise the location of the NS5-brane intersection from the D6$_1$-brane worldvolume, we have a non-trivial 3-cycle on which there is one unit of $H_3$ flux, leading to a Freed-Witten inconsistency \cite{Freed:1999vc,Maldacena:2001xj}. This forces each of the D6$_1$-branes to emit one D4-brane, spanning 012 times the radial direction in 3456 times one direction away from the D6$_1$-brane worldvolume.

Conversely, the flux created by the D6$_1$-branes implies a Freed-Witten inconsistency on the NS5-brane, as follows. The intersection of the D6$_1$-brane with the NS5-brane is codimension 3 in the NS5-brane worldvolume, hence an $\IS^2$ surrounding the D6$_1$-brane in the NS5-brane worldvolume (namely the angular part in the $\IR^3$ spanned by 789) supports a RR 2-form field strength flux
\beqa
\int_{\IS^2} F_2=N_1 \, .
\eeqa
This forces the NS5-brane to emit $N_1$ D4-branes, spanning 012 times the radial direction in 789 times a direction transverse to the NS5-brane worldvolume. 

\begin{figure}[htb]
\begin{center}
\includegraphics[scale=.3]{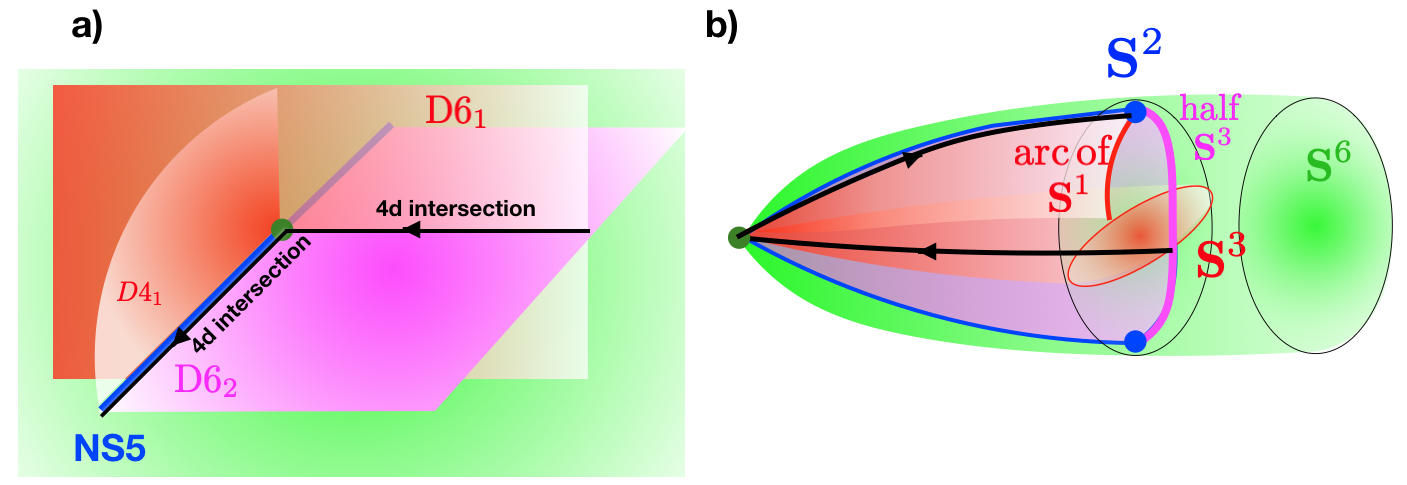}
\caption{\small Stack of infinite D6$_1$-branes intersecting a stack of semi-infinite D6$_2$-branes (ending on an NS5-brane) over 4d half-space in flat 10d space, with the D4-branes suspended between the infinite D6$_1$-branes and the NS5-brane. a) The view in flat space. b) The cone perspective. Comparing with Figure \ref{fig:half}, the addition of the D4-branes lead to additional intersections supporting 4d chiral fermions in the bifundamental representation, so that this spectrum is now defined on an infinite line.}.
\label{fig:thed4s}
\end{center}
\end{figure}

Overall, and keeping track of the orientations, we end up with $N_1$ D4-branes stretching between the D6$_1$-branes and the NS5-brane, see Figure \ref{fig:thed4s}a. Note that the D4-branes indeed span a radial direction away from the intersection of the D6$_1$-branes and the NS5-branes, both on the worldvolume of the D6$_1$-branes and of the NS5-brane, and one direction transverse to the D6$_1$-brane worldvolume and one direction transverse to the NS5-brane worldvolume. In the cone construction, the wedge spanned by the D4-branes is a cone over an arc of $\IS^1$stretching between (a point in) the $\IS^3$ of the D6$_1$ and (a point in) the $\IS^2$ of the NS5-brane, see Figure \ref{fig:thed4s}b.

%%%%%%%%%%%
%%%%%%%%%%%

Consider now the implications of the D4-branes for the spectrum of the theory, starting in the flat space configuration, see Figure \ref{fig:thed4s}a. A crucial observation is that when the D4-brane ends on the D6-brane, their gauge groups are identified. This is analogous to the familiar statement that D3-branes ending on D5-branes have Dirichlet boundary conditions for the vector multiplets \cite{Hanany:1996ie}. To emphasize this, in Figure \ref{fig:thed4s} we have labeled the D4-branes with a subindex 1, and we have colored them in red, just like the D6$_1$-branes. This also agrees well with the fact that there is one D4-brane per D6$_1$-brane, so the two stacks collectively carry both a single $U(N_1)$. 

Hence, in the sector of open strings stretching between the D4-branes and the D6$_2$-branes, we obtain matter in the bifundamental of the $SU(N_2)$ on the D6$_2$-branes and the $SU(N_1)$ on the D6$_1$/D4$_1$-branes. We note that the spectrum at the intersection of D4-branes and D6-branes ending on the same NS5-brane was shown in \cite{Hanany:1997sa} to indeed correspond to this kind of 4d chiral bifundamental $(\fund_1,\antifund_2)$. This is precisely the content we had on the 4d intersection of the D6$_1$- and D6$_2$-branes, so we have indicated it Figure \ref{fig:thed4s} with the same black arrow line. Note that the new black line from the D6$_2$-D4$_1$ intersection continues the formerly semi-infinite  black line from the D6$_1$-D6$_2$ intersection. This implements the analogue of the fate of worldsheet fermions in the open heterotic string in section \ref{sec:open-het}: The 4d chiral fermion at the D6$_1$-D6$_2$  intersection reaches the boundary of its support, but it is carried away by some additional degrees of freedom, in this case the D6$_2$-D4$_1$ intersection.

This picture makes the anomaly inflow consistent. The inflow from the D6$_2$-brane bulk into the intersection with the D6$_1$-brane continues as an inflow towards the intersection with the D4$_1$-brane. Similarly, the inflow from the D6$_1$-brane bulk into the intersection with the D6$_2$-brane turns into an inflow from the D4$_1$-brane into the intersection with the D6$_2$-brane. This is all consistent with the interpretation of an inflow for a continuous D6$_2$-D6$_1$/D4$_1$ intersection.

Turning now to the cone construction and keeping track of the orientations, see Figure \ref{fig:thed4s}b, the compactification of the 10d theory on $\IS^6$ now produces a 4d theory with one chiral fermion in the $(\fund_1,\antifund_2)$, from the D6$_1$-D6$_2$ intersection, and one in the $(\antifund_1,\fund_2)$, from the D6$_2$-D4$_1$ intersection. The complete spectrum is therefore non-chiral, again in analogy with the cone construction for the open heterotic string in section \ref{sec:cone-open-het}.

The  cone construction hence describes an interesting dynamical cobordism in a theory with a non-trivial sector of 4d fermions, but ultimately a non-chiral one. In appendix \ref{app:more-intersecting} we quickly describe other variants with a similar set of ingredients\footnote{\label{foot-D8} It is also possible to define semi-infinite D6-branes by allowing them to end on D8-branes. In general, such configurations eventually fail to produce boundary configurations for chiral theories because the gauge groups on D6-branes are linked to those of the D8-brane on which they end. Hence, a configuration of intersecting D6-branes ending on a D8-brane fails to produce chirality because all gauge factors become identified, collapsing the bifundamental fermion onto some non-chiral representation.}, again leading to non-chiral theories upon the cone construction. These examples illustrate that getting chirality in the cone construction is thus highly non-trivial. In the next section we will identify a key property underlying the non-chirality in these examples, and will overcome it and obtain a new large class of constructions of boundary configurations for genuine 4d chiral theories.

\subsection{Chiral cones from Branes at Singularities}
\label{sec:cone-singus}

A main reason why the above constructions ultimately lead to non-chiral theories upon the cone construction is that the gauge groups extend in more than one dimensions away from the tip of the cone (i.e. the boundary of the 4d fermion). This implies that the two copies of chiral fermions arising over the base of the cone are charged under the same gauge factors and lead to a non-chiral configuration. 

Fortunately, there are several ways to obtain 4d chiral fermions from D-branes  (see \cite{Ibanez:2012zz} for a review). In addition to intersecting D6-branes (or their mirror realization, magnetized-branes), it can be achieved using D3-branes at singularities \cite{Douglas:1996sw,Lawrence:1998ja,Ibanez:1998xn,Hanany:1998it} (see \cite{Aldazabal:2000sa} for model building applications and \cite{Ibanez:2012zz} for review). Hence, it is natural to resolve the above problems by using 4d chiral fermions from D-branes of lower dimension, specifically D3-branes at singularities, as we explore in this section. Incidentally, the resulting construction bears some analogies with the 6d setup discussed in section \ref{sec:open-m5}.

\subsubsection{Example: Cone over the dP$_0$ theory}
\label{sec:dp0-example}

For concreteness, we illustrate the main construction in an explicit example, known as the dP$_0$ theory, leaving the general construction for section \ref{sec:examples-singus}. To build the dP$_0$ theory, consider a stack of D3-branes at the tip of a $\IC^3/\IZ_3$ orbifold singularity, with the generator $\theta\in\IZ_3$ acting on the $\IC^3$ coordinates as 
\begin{equation}
\theta:(z_1,z_2,z_3)\to (e^{2\pi i/3}z_1,e^{2\pi i/3}z_2, e^{-4\pi i/3}z_3)\,.
\label{z3-generator}
\end{equation}
We choose the orbifold action on the Chan-Paton indices in $N$ copies of the regular representation
\beqa
\gamma_{\theta}={\rm diag}({\bf 1}_N,e^{2\pi i/3}{\bf 1}_N, e^{4\pi i/3}{\bf 1}_N)\, .
\eeqa 
The resulting 4d $\NN=1$ gauge theory on the D3-branes\footnote{We are removing the $U(1)$ factors as they are made massive by St\"uckelberg couplings, which in this case are required for the 4d Green-Schwarz mechanism \cite{Ibanez:1998qp}.} has gauge group and chiral multiplet content given by
\beqa 
&SU(N)_0\times SU(N)_1\times SU(N)_2 &\nonumber \\
&3\,[\, (\fund_0,\antifund_1,{\bf 1})\,+\,({\bf 1},\fund_1,\antifund_2)\,+\, (\antifund_0,{\bf 1},\fund_2)\,]\, ,&
%&W= \epsilon_{ijk} X_{01}^iX_{12}^j X_{20}^k&
\label{dp0-first}
\eeqa 
and there is a cubic superpotential which we skip for the moment. Note that the cubic anomalies cancel, as expected as a consequence of twisted RR tadpole cancellation \cite{Aldazabal:1999nu}, automatically satisfied for the regular representation.

Let us now perform a cone construction using the above configuration. Consider $\IC^3/\IZ_3\times \IR$, where the $\IR$ corresponds to one of the directions along the D3-branes, say $x^3$. The full 7d space can be regarded as $\IR^7$ modded out by a $\IZ_3$ quotient acting on the first 6 real coordinates and leaving $x^3$ invariant. This can be regarded as a real cone over a 6d base $\IS^6/\IZ_3$, where the $\IS^6$ (of radius $R$) is described as
\beqa
|z_1|^2 +|z_2|^2+|z_3|^2+(x^3)^2=R^2\, ,
\eeqa 
and the generator of $\IZ_3$ acts on it as in (\ref{z3-generator}). Hence, we can regard the configuration as a 4d compactification of type IIB theory on $\IS^6/\IZ_3$, in which the size of the internal space runs along a 4d spacetime coordinate. Namely, it corresponds to a dynamical cobordism in the spirit in \cite{Buratti:2021fiv,Angius:2022aeq,Blumenhagen:2022mqw,Blumenhagen:2023abk}, a connection which we make more explicit in a related class of constructions in section \ref{sec:dyn-cob-d3s}.

Let us now give some more details about the content of this 4d theory. On the $\IS^6$ there are two fixed points corresponding to $z_i=0$, $x^3=\pm R$, at each of which there is a system of D3-branes at a local $\IC^3/\IZ_3$ singularity, leading to the 4d chiral spectrum (\ref{dp0-first}). Due to the different orientation (since increasing the radial coordinate corresponds to increasing or decreasing $x^3$ at the two points, respectively), what we have is a system of D3-branes and anti-D3-branes. The 4d chiral spectrum arising at this point is thus again given by a copy of (\ref{dp0-first}), and a second copy with 4d fermions of the opposite chirality \cite{Aldazabal:2000sa} (recall that we abuse language with the use of susy jargon, c.f. footnote \ref{foot:non-susy}). In contrast with the previous section, the gauge groups now arising at the two points are two independent sets, and therefore the fermions at the two points are charged in conjugate bifundamentals but under {\em different} sets. Hence, the resulting 4d theory is chiral, and the cone construction, regarded as dynamical cobordism in the 4d theory, provides a boundary configuration for a genuinely 4d chiral theory.

The fact that the two stacks correspond to D3-brane / anti-D3-brane pairs suggests that the dynamical mechanism that explains the gapping of the 4d chiral degrees of freedom corresponds to a brane-antibrane annihilation process. As expected, this is beyond the regime admitting a weakly coupled description, or even a field theory description, since open string tachyon condensation can be properly described only in string field theory. Let us note that however, in analogy with the 6d example in section \ref{sec:open-m5}, it is possible to provide a simpler effective description of the resulting boundary conditions. Indeed, the boundary conditions at the tip of the cone amount to an exchange of left- and right chiralities, with a simultaneous exchange of the two singularities and their corresponding gauge sectors, i.e. a $\IZ_2$ outer automorphism symmetry of the theory, which is a symmetry  of the underlying geometry of the base of the cone. Overall, the mechanism is a close cousin of that in the  bubble of nothing in \cite{Fabinger:2000jd}, when regarded from the 10d perspective, in which identical sectors preserving different supersymmetries annihilate against each other at the ETW brane.

Let us also mention that, although the 4d theory under discussion is highly non-supersymmetric, the final running solution describing the dynamical cobordism is supersymmetric, as it secretly corresponds to the system of D3-branes at an orbifold of flat space. The fact that dynamical cobordism solutions may enjoy more supersymmetry than the effective theory is familiar from several other examples, see e.g. \cite{Buratti:2021yia}.

\subsubsection{Generalization}
\label{sec:examples-singus}

The above example admits a straightforward generalization to a large class of configurations. As discussed in \cite{Klebanov:1998hh,Morrison:1998cs}, there are large classes of CY3 singularities $\IX_6$ built as cones over 5d geometries $\IY_5$, for which the corresponding gauge theory on D3-brane probes can be identified. In particular, for toric singularities $\IX_6$ there is a specific dictionary via dimer diagrams (a.k.a. brane tilings) \cite{Hanany:2005ve,Franco:2005rj,Feng:2005gw,GarciaEtxebarria:2006aq}(see \cite{Kennaway:2007tq} for a review), allowing to read out the gauge theory from geometric data, and vice versa, which has been extensively exploited in holography \cite{Franco:2005rj} (see also e.g. \cite{Franco:2005fd,Berenstein:2005xa,Franco:2005zu,Bertolini:2005di}) and model building, see e.g.\cite{Cascales:2005rj,Garcia-Etxebarria:2006lri}. We will discuss this specific dictionary in section \ref{sec:mirror-d3s}, but it is not necessary  in this section, where we keep the discussion general.

We hence consider a system of D3-branes at a (not necessarily toric) CY3 singularity $\IX_6$, leading to a 4d chiral gauge theory with group $G$ (a product of unitary factors) and 4d chiral fermions in a representation ${\cal R}$. Let us now consider the 7d geometry $ \IX_7= \IX_6\times \IR$, with $\IR$ parameterized by one of the coordinates along the D3-branes, say $x^3$. Let us write the metric as
\beqa
ds_7^2= (dx^3)^2 + dr'^2 + r'^2 ds_{\IY_5}^2\, ,
\label{cone-7d-one}
\eeqa  
with $r'>0$. Defining polar coordinates in the $(r',x^3)$ 2-plane, i.e. $r'=r\cos\theta$, $x^3=r\sin\theta$, we have
\beqa
ds_7^2= dr^2 + r^2 (d\theta^2+ \cos^2\theta  ds_{\IY_5}^2 )\, ,
\label{cone-7d-two}
\eeqa
which describes the 7d geometry as a real cone, with radial coordinate $r$ and base geometry $\IY_6$, given by the suspension of $\IY_5$, i.e. the fibration of $\IY_5$ over a segment, parametrized by $\theta\in[-\pi/2,\pi/2]$, with the fiber collapsed to a point over the two endpoints. The locus $r'=0$ and arbitrary $x^3$ is a real line of singularities locally identical to $\IX_6$, located at $\theta=\pm \pi/2$ and arbitrary $r$ in polar coordinates in $\IX_7$. 

Hence, we have a 4d theory (in the directions 012 and $r$) obtained by compactification of type IIB theory on $\IY_6$, with D3-branes and antibranes located at the points $\theta=\pm \pi/2$, respectively, in the internal space. There is a 4d gauge group $G\times G$, and 4d chiral fermions  in the representation $({\cal R},{\bf 1})+({\bf 1}, {\cal R})$. This leads to a large class of 4d chiral theories for which the above construction produces boundary configurations described as dynamical cobordisms to nothing.

It is interesting to point out that the boundary condition effectively exploits a combination of chirality flip and the $\IZ_2$ outer automorphism exchanging the two gauge theories. This resembles the behaviour in the bubble of nothing in \cite{Fabinger:2000jd}, when regarded from the 10d perspective (it is also reminiscent of the folding trick used to define boundary states in 2d theories).

Despite its appeal, the explicit presence of branes and antibranes in the configuration, equivalently of two copies of the gauge sector (with opposite chiralities) in the 4d theory, makes this construction less enticing. In the next section we will present a variation, which improves on this respect.

\subsection{Boundaries from $\IZ_2$ quotients}
\label{sec:z2-quotients}

In this section we build on the construction in the previous section to obtain new classes of boundary configurations for chiral 4d theories. They are inspired in $\IZ_2$ quotients used in the construction of barely $G_2$ holonomy spaces \cite{Joyce:1996,Harvey:1999as}, which we review next.

\subsubsection{D6-branes at $G_2$ holonomy 7d geometries}
\label{sec:g2}

Given a CY3 $\IX_6$, which can be compact or not in this general discussion, we consider the quotient $\IX_7=(\IX_6\times \IR)/\IZ_2$, with the generator $R\in\IZ_2$ acting as $x^3\to -x^3$ on the coordinate parametrizing $\IR$, and as an antiholomorphic action on $\IX_6$. Specifically, the action on the K\"ahler form $J$ and holomorphic 3-form $\Omega$ are
\beqa
R(J)=-J\quad , \quad R(\Omega)={\ov {\Omega}}\, .
\label{antiholomorphi-involution}
\eeqa
The resulting $\IX_7$ is a 7d barely $G_2$ holonomy space with covariantly constant 3-form
\beqa 
\varphi_3=Jdx^3 + {\rm Re}\, (\Omega)\, ,
\eeqa 
which is clearly invariant under the action of $R$. The term {\em barely} reflects the fact that the actual holonomy is an $SU(3)\ltimes \IZ_2$ subgroup of $G_2$. This kind of construction has been exploited in the M-theory lifts of type IIA configurations in \cite{Kachru:2001je}.

We are interested to consider type IIA models on $\IX_6$ supporting 4d chiral fermions. So we consider stacks of $N_a$ D6-branes wrapped on special lagrangian 3-cycles $\Pi_a$ of $\IX_6$, corresponding to intersecting brane models \cite{Blumenhagen:2000wh,Aldazabal:2000dg,Aldazabal:2000cn,Cvetic:2001nr,Cvetic:2001tj}, see \cite{Blumenhagen:2006ci,Ibanez:2012zz} for review. Note that, although in the compact setup these models are non-supersymmetric unless O6-planes are introduced \cite{Cvetic:2001nr,Cvetic:2001tj}, for non-compact CY threefolds the additional freedom in RR tadpole cancellation allows for supersymmetric models with D6-branes wrapped on compact 3-cycles \cite{Uranga:2002pg}, so we focus on the latter setup. The supersymmetric 3-cycles are defined by the condition that they satisfy the special lagrangian conditions
\beqa
J\big|_{\Pi_a}=0\quad ,\quad {\rm Im}\, (\Omega)\big|_{\Pi_a}=0\, .
\eeqa
Equivalently, the 3-cycles are calibrated with respect to the 3-form ${\rm Re}\, (\Omega)$ \cite{Becker:1995kb,Becker:1996ay}.

As is familiar, the 4d $\NN=1$ spectrum contains, an $SU(N_a)$ gauge groups on each D6-brane stack (the $U(1)$ factors are generically massive due to St\"uckelberg couplings), and their intersections lead to a net number $I_{ab}=[\Pi_a]\cdot[\Pi_b]$ of chiral multiplets in the bifundamental $(\fund_a,\antifund_b)$ representation.

It is now easy to check that the above supersymmetric 3-cycle conditions in $\IX_6$ are invariant under $\IZ_2$ action $R$, so each individual 3-cycle $\Pi_a$ is either invariant under $R$, or exchanged with another supersymmetric 3-cycle, denoted by $\Pi_{a'}$. Starting with a $\IZ_2$ invariant set of D6-branes wrapped on such supersymmetric 3-cycles, they descend to D6-branes wrapped on supersymmetric coassociative 4-cycles in the $G_2$ geometry $\IX_7$. Specifically, namely they are calibrated with respect to the 4-form $*_{7d}\varphi_3$. 

Given one such D6-brane configuration, there is a spectrum of 4d chiral fermions localized on real lines (parametrized by $x^3$) in $\IX_7$, as follows. In the covering space of the $\IZ_2$ quotient, we have the 4d chiral theory described above. The effect of the $\IZ_2$ action on this theory is as follows: a generic 3-cycle $\Pi_a$ in $\IX_6$, at a location $x^3$ in $\IR$, is mapped to the image 3-cycle $\Pi_{a'}$ at the location $-x^3$ in $\IR$, and such that the fundamental $\fund_a$ is mapped to the {\em same} representation $\fund_{a'}$, because this is an orbifold, rather than an orientifold, projection. Hence, at a point $x^3$ in $\IR$, the effect of the $\IZ_2$ projection is not felt locally, and the local 4d spectrum we obtain is as described in the previous paragraphs. However, at the image point $-x^3$ in the double cover, the degrees of freedom are not independent, but are a mere image of them.  In particular notice that remarkably a 4d chiral multiplet $\Phi_{ab}$ in a bifundamental $(\fund_a,\antifund_b)$ of $SU(N_a)\times SU(N_b)$ at a location $x^3$ is related to a bifundamental chiral multiplet $\Phi_{b'a'}$ in the $(\antifund_{a'},\fund_{b'})$ of the image group\footnote{For groups mapped to themselves under the $\IZ_2$ action, we postpone the discussion to the explicit examples in later sections.} $SU(N_{a'})\times SU(N_{b'})$ at the location $-x^3$, namely the $(\antifund_{a},\fund_{b})$ of $SU(N_a)\times SU(N_b)$. The two sets of degrees of freedom seem to be in different (in particular conjugate) representations of the gauge group, which would make the $\IZ_2$ identification impossible. However, we should notice that the $\IZ_2$ generator $R$ acts on $x^3$ as a parity operation, thus flipping the 4d chirality of the corresponding fermion, and this precisely compensates the conjugation of the gauge representation. In other words, the identification by the full orbifold action is $\Phi_{ab}\leftrightarrow {\ov {{\Phi}_{b'a'}}}$, which implies the chirality flip and the conjugation of quantum numbers. Hence the action of the $\IZ_2$ quotient is consistent and defines a consistent identification of degrees of freedom in the spectrum. In short, we get one copy of the 4d chiral gauge theory in the $\IZ_2$ quotient.
This construction holds the key to the removal of the doubling of degrees of freedom encountered in section \ref{sec:cone-singus}. 

We would now like to obtain models of 4d chiral theories by taking local models of intersecting D6-branes on a non-compact Calabi-Yau $\IX_6$, and carry out a 7d cone construction involving the extra $x^3$ coordinate. However, obtaining a global 7d cone is possible only if $\IX_6$ is a cone itself. We present one particular explicit example in Appendix \ref{app:g2-cone-example}, based on a model in \cite{Acharya:2003ii}. However, such explicit examples are scarce, due to the familiar difficulties to build special lagrangian 3-cycles in general CY3s. Hence, in the following section, we instead turn to the implementation of the above construction in systems of D3-branes at singularities, where the cone structure is built in from the beginning, and so it leads to a large class of explicit examples.

\subsubsection{The type IIB picture}
\label{sec:mirror-d3s}

A large class of models of supersymmetric local intersecting D6-brane models \cite{Uranga:2002pg}, namely D6-branes on compact 3-cycles on non-compact CY3 geometries, can be obtained as the mirror of the systems of D3-branes at toric CY3 singularities, mentioned in section \ref{sec:examples-singus}, which are efficiently studied using dimer diagrams (a.k.a. brane tilings) \cite{Hanany:2005ve,Franco:2005rj,Feng:2005gw,GarciaEtxebarria:2006aq}(see \cite{Kennaway:2007tq} for a review). In particular, the mirror map between D3-branes at toric CY3 singularities and local intersecting D6-brane models can be carried out systematically via the explicit map in \cite{Feng:2005gw}. This map allows to perform a construction similar to that in the previous section, but in systems of D3-branes at conical CY3 singularities. This setup will be best suited to subsequently perform a Chiral Cone construction and lead to boundary configurations for large classes of 4d chiral theories.

Consider a system of D3-branes at a CY3 toric singulary $\IX_6$. We momentarily focus on regular D3-brane systems (i.e. all gauge factors have the same rank), although in later discussions we will allow for fractional branes (i.e. anomaly-free rank assignments for the different nodes). As explained, the 4d $\NN=1$ gauge theory is efficiently encoded in dimer diagrams, as we will make explicit in concrete examples in section \ref{sec:examples}.

The mirror geometry ${\tilde{\IX}_6}$ is constructed as a base $\IC$ parametrized by a coordinate $z$, over which we fiber a $\IC^*$, with the fiber degenerating at $z=0$, and a Riemann surface $\Sigma$, with various 1-cycles $C_i$ degenerating at various points $z_i$ on the base, as explained later. Namely, the geometry is described as
\beqa
uv=z\quad , \quad P(w_1,w_2)=z\, ,
\label{mirror-geometry}
\eeqa
where $u,v$ parametrize the $\IC^*$ fiber, an the second equation describes $\Sigma$, with $P(w_1,w_2)$ the Newton polynomial of the toric geometry. The compact special lagrangian 3-cycles wrapped by the D6-branes mirror to the D3-branes in some node $i$ of the quiver are  obtained as follows: one takes a segment on the base joining $z=0$ and the degeneration point $z_i$ of some 1-cycle $C_i\subset\Sigma$, and fibers the $\IS^1\subset\IC^*$ times $C_i\subset\Sigma$. The result is a set of topological 3-spheres $\IS^3_i$, shown in \cite{Feng:2005gw} to lead to the intersections and worldsheet instantons to yield the spectrum and interactions of the original D3-brane theory. The mirror geometry is hence basically controlled by the geometry of the fiber $\Sigma$ and its set of degenerating 1-cycles. The construction of this mirror Riemann surface and the 1-cycles wrapped by the D6-branes will be carried out in explicit examples using the procedure in \cite{Feng:2005gw}.

One may thus carry out the $({\tilde X}_6\times \IR)/\IZ_2$ quotient on the type IIA system with intersecting D6-branes, but this does not allow for a cone construction because ${\tilde X}_6$ is not conical. So the strategy is to return to the original picture of D3-branes at the cone $\IX_6$, and consider now the space $(\IX_6\times \IR)/\IZ_2$, where the generator $R\in\IZ_2$ acts as $x^3\to -x^3$ in $\IR$ and as a $\IZ_2$ involution on $\IX_6$ (also denoted by $R$, with abuse of language) corresponding to the antiholomorphic action in the type IIA mirror geometry ${\tilde{\IX}}_6$, just mentioned. The action of $R$ on the gauge theory can be read as an action on the dimer diagram, via the explicit mirror map. As we will show in explicit examples, it also corresponds to an antiholomorphic $\IZ_2$ action on $\IX_6$, so that the full quotient preserves half of the supersymmetries. The resulting 7d space thus has $G_2$ holonomy, and presumably corresponds to the mirror of the type IIA $G_2$ manifold in the sense of \cite{Acharya:1997rh}. The fixed loci are orbifold 5-planes, wrapped on special lagrangian 3-cycles in $\IX_6$ and sitting at $x^3=0$, where they define a boundary configuration in the quotient\footnote{These orbifold 5-planes are S-dual to configurations of O5$^-$-planes with D5-branes on top \cite{Sen:1996na}, which have been exploited to define boundary configurations in compactifications \cite{Friedrich:2023tid} and in holographic setups \cite{GarciaEtxebarria:2024jfv,Huertas:2024mvy}. It would be interesting to explore further connections with these setups.}. 

Compared with the models in section \ref{sec:cone-singus}, the effective boundary condition for the 4d theory involves a chirality flip and a $\IZ_2$ action on the gauge theory, such that the combined action is a symmetry. The $\IZ_2$ action on the gauge theory can in general be a combination of inner and outer automorphisms of the different gauge factors.

The different $\IZ_2$ actions on dimer diagrams were studied, in the context of orientifold quotients\footnote{The reason why orientifold actions appear as the relevant quotients in our context is because the orbifold includes a parity flip in the direction $x^3$, which acts on the fermions by conjugation of quantum numbers (equivalently, by a chirality flip, as befits to the definition of a boundary condition), an operation which, for actions preserving Poincar\'e invariance, arises only in orientifold quotients.}, in \cite{Franco:2007ii}, and correspond to reflections leaving fixed points or fixed lines in the dimer diagram.  As will be clear from the examples in section \ref{sec:examples}, the antiholomorphic involutions in the type IIA mirror correspond to actions with fixed lines. This allows to efficiently describe the effect of the $\IZ_2$ involution $R$ in large classes of dimer gauge theories. We will present specific examples in later sections. 

The construction in the IIB side allows for a cone construction because $\IX_6$ is a cone, hence so is $(\IX_6\times \IR)/\IZ_2$. As in section \ref{sec:examples-singus}, we take the  metric in $\IX_6\times \IR$ 
\beqa
ds_7^2= (dx^3)^2 + dr'^2 + r'^2 ds_{\IY_5}^2\, ,
\eeqa  
where $\IX_6$ is written as a real cone over the 5d base $\IY_5$. Using polar coordinates in the $(r',x^3)$ 2-plane, i.e. $r'=r\cos\theta$, $x^3=r\sin\theta$, we have
\beqa
ds_7^2= dr^2 + r^2 ( d\theta^2+\cos^2\theta  ds_{\IY_5}^2 )\, ,
\label{cone-7d}
\eeqa
which describes the 7d geometry as a real cone over the base geometry $\IY_6$, given by the suspension of $\IY_5$. The real line of singularities locally identical to $\IX_6$ is the locus $r'=0$ and arbitrary $x^3$, equivalently  $\theta=\pm \pi/2$ and arbitrary $r$. 

Performing the $\IZ_2$ quotient, the coordinate $r$ is invariant, while we have a non-trivial quotient $\theta\to -\theta$. This means that the quotient geometry is of the kind (\ref{cone-7d}), with the restriction $\theta\in[0,\pi/2)$. We thus have a real cone over the 6d space given by a quotient of the suspension of $\IY_5$. The locus corresponding to the singularity $\IX_6$ is now given by just $\theta=\pi/2$. 

This implies that, in the 4d theory obtained by reducing on the 6d base $\IY_6$ of the cone, there is a single copy of the 4d $\NN=1$ gauge theory corresponding to the systems of D3-branes at the tip of a local $\IX_6$ singularity. The cone construction for $\IX_7$ is hence a Chiral Cone construction, providing a boundary configuration for this 4d $\NN=1$ gauge theory, coupled to gravity. The resulting 4d configuration describes a running solution in which the scalar corresponding to the $\IY_6$ size varies along the direction $r$. It corresponds to a dynamical cobordism in which at a finite spacetime distance point $r=0$ the scalar blows up, the internal space shrinks to zero size and spacetime ends. This solution is discussed in more detail in section \ref{sec:dyn-cob-d3s}.

As already mentioned, one should regard the configuration as a local description of a possibly more involved global solution, which moderates the asymptotic growth of $\IY_6$ e.g. to a constant size, in analogy with Witten's bubble of nothing.

\subsubsection{Examples}
\label{sec:examples}

{\bf The dP$_0$ theory}

\begin{figure}[htb]
\begin{center}
\includegraphics[scale=.3]{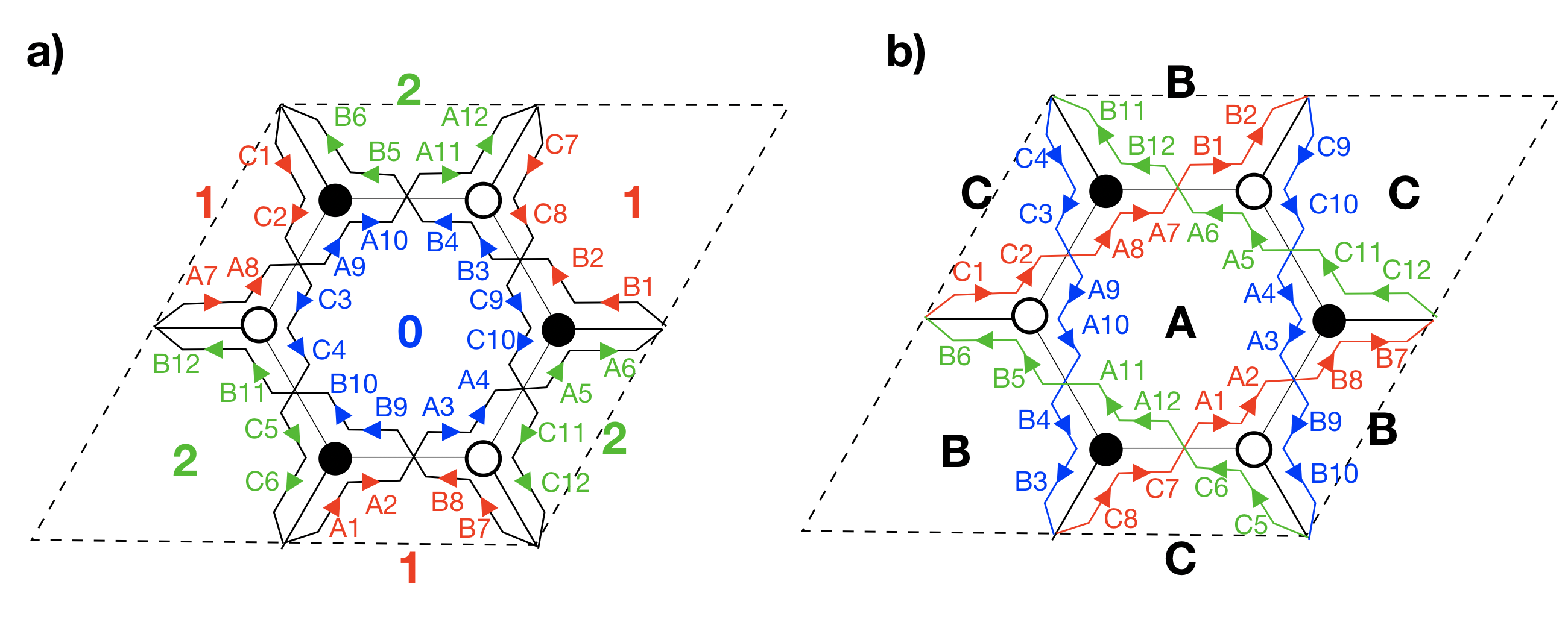}
\caption{\small a) The dimer diagram for the dP$_0$ theory, with its set of zig-zag paths. b) The mirror Riemann surface obtained via the untwisting procedure in \cite{Feng:2005gw}.}
\label{fig:mirror-dp0}
\end{center}
\end{figure}

\begin{figure}[htb]
\begin{center}
\includegraphics[scale=.3]{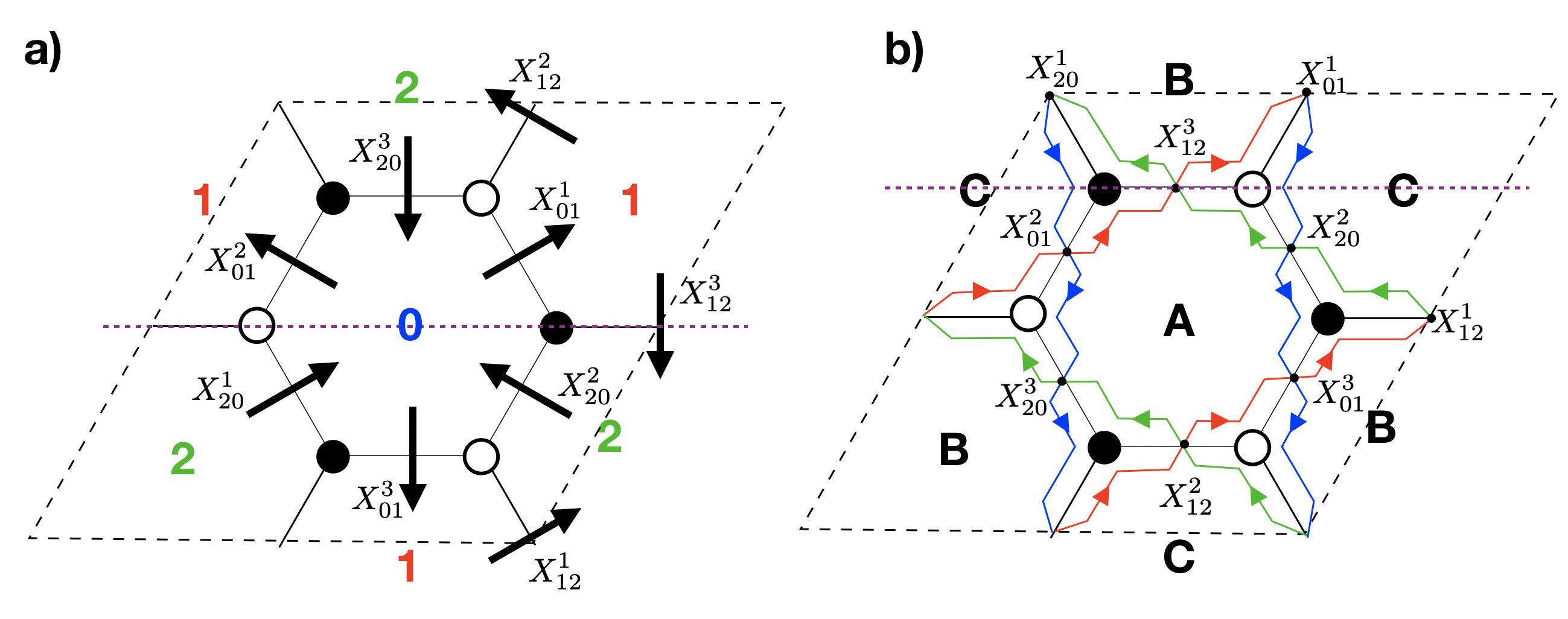}
\caption{\small a) The dimer diagram for the dP$_0$ theory, with arrows indicating the  bifundamental chiral multiplets. We also display the $\IZ_2$ involution as a reflection with respect to the fixed line, indicated as a dashed violet line. b) The mirror Riemann surface with the D6-brane cycles in colors, and with the location of the bifundamental chiral multiplets at their intersections. The action of the $\IZ_2$ involution matching that in Figure (a) corresponds to an antiholomorphic action, of the kind studied in section \ref{sec:g2}.}
\label{fig:z2-action-dp0}
\end{center}
\end{figure}

In order to illustrate the above construction in practice, we consider a few illustrative examples. Let us consider the dP$_0$ theory, which is obtained from D3-branes at a $\IC^3/\IZ_3$ singularity (i.e. a complex cone over dP$_0=\IP_2$), already appeared in section \ref{sec:dp0-example}. Because this is an orbifold of flat space, the gauge theory can be determined using standard worldsheet techniques for the open string sectors. The geometry is toric, hence the theory has a dimer diagram description shown in Figures \ref{fig:mirror-dp0}a, \ref{fig:z2-action-dp0}a. The gauge theory is given by
\beqa 
&SU(N)_0\times SU(N)_1\times SU(N)_2 &\nonumber \\
&3\,[\, (\fund_0,\antifund_1,{\bf 1})\,+\,({\bf 1},\fund_1,\antifund_2)\,+\, (\antifund_0,{\bf 1},\fund_2)\,]&\nonumber\\
&W= \epsilon_{ijk} X_{01}^iX_{12}^j X_{20}^k\, ,&
\label{dp0}
\eeqa 
where the bifundamental fields in the superpotential have subindices indicating the gauge representation (in a hopefully self-explanatory way) and a superindex labelling the three copies of each field (and which correspond to the three complex coordinates $(z^1,z^2,z^3)\in\IC^3$). Also, the trace in the superpotential is implicit here and in what follows. For completeness, we display in Figure \ref{fig:mirror-dp0}b the mirror Riemann surface obtained via the untwisting procedure in \cite{Feng:2005gw}.

%%%%%%%%%%%
%%%%%%%%%%%

In Figure \ref{fig:z2-action-dp0}a we have indicated the $\IZ_2$ involution required to build the 7d geometry $\IX_7$. Specifically, the gauge group $SU(N)_0$ is mapped to itself, while $SU(N)_1$ and $SU(N)_2$ are exchanged. On the bifundamental matter, we have the action 
\beqa
R: & X^1_{01}\leftrightarrow {\ov X}^2_{20}\quad , \quad X^1_{20}\leftrightarrow {\ov X}^2_{01} \nonumber\\
& X^1_{12} \leftrightarrow {\ov X}^2_{12} \quad , \quad X^3_{01}\leftrightarrow  {\ov X}^3_{20}\, ,
\label{z2-fields-dp0}
\eeqa 
with $X^3_{12}$ being mapped to its conjugate. As mentioned above, it is easy to check that in the mirror geometry, this action corresponds to an antiholomorphic involution of the kind discussed in section \ref{sec:g2}. The action on the coordinate $z$ on the base and the coordinates $u,v$ in the $\IC^*$ fiber in (\ref{mirror-geometry}) is $z\to {\ov z}$, $u,v\to {\ov u},{\ov v}$, while the action to the Riemann surface in shown in Figure \ref{fig:z2-action-dp0}b. The action is thus antiholomorphic, and acts as (\ref{antiholomorphi-involution}) on the mirror geometry ${\tilde{\IX}_6}$.

%%%%%%%%%%%
%%%%%%%%%%%
%

The action (\ref{z2-fields-dp0}) on the fields can be translated into a geometric action on the type IIB $\IC^3/\IZ_3$ geometry. Recall that the coordinates of the transverse space to the D3-branes can be constructed as the gauge invariant mesons of the quiver theory. For $\IC^3/\IZ_3$ we specifically have
\beqa
z^1=X^1_{01}X^1_{12}X^1_{20}\quad ,\quad z^2=X^2_{01}X^2_{12}X^2_{20}\quad ,\quad z^3=X^3_{01}X^3_{12}X^3_{20}\, ,
\eeqa 
modulo F-term relations. Using (\ref{z2-fields-dp0}), this corresponds to 
\beqa
R: (z_1,z_2,z_3)\,\to\,({\ov z}_2,{\ov z}_1,{\ov z}_3)\, .
\eeqa 
As anticipated, the antiholomorphic action on the CY threefold $ \IC^3/\IZ_3$, together with $x_3 \mapsto - x_3$, defines a $7d$ $G_2$ orbifold, which thus preserves half of the supersymmetries.  Following the above Chiral Cone construction, we have a boundary configuration for a 4d $\NN=1$ chiral theory with spectrum (\ref{dp0}) coupled to gravity.

\medskip

{\bf The $\IF_0$  theory}

\begin{figure}[htb]
\begin{center}
\includegraphics[scale=.28]{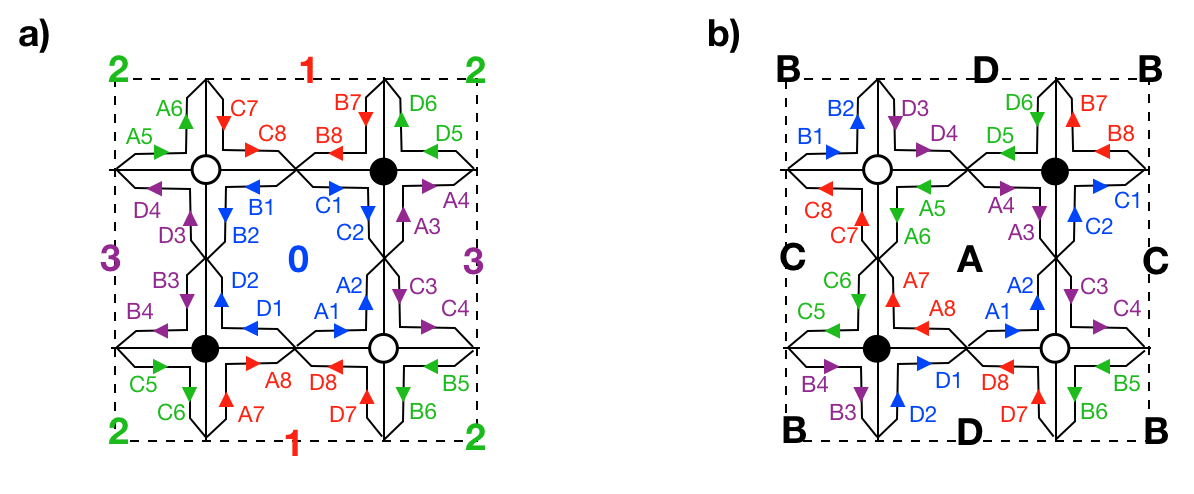}
\caption{\small a) The dimer diagram for the $\IF_0$ theory, with its set of zig-zag paths. b) The mirror Riemann surface obtained via the untwisting procedure in \cite{Feng:2005gw}.}
\label{fig:mirror-f0}
\end{center}
\end{figure}

\begin{figure}[htb]
\begin{center}
\includegraphics[scale=.3]{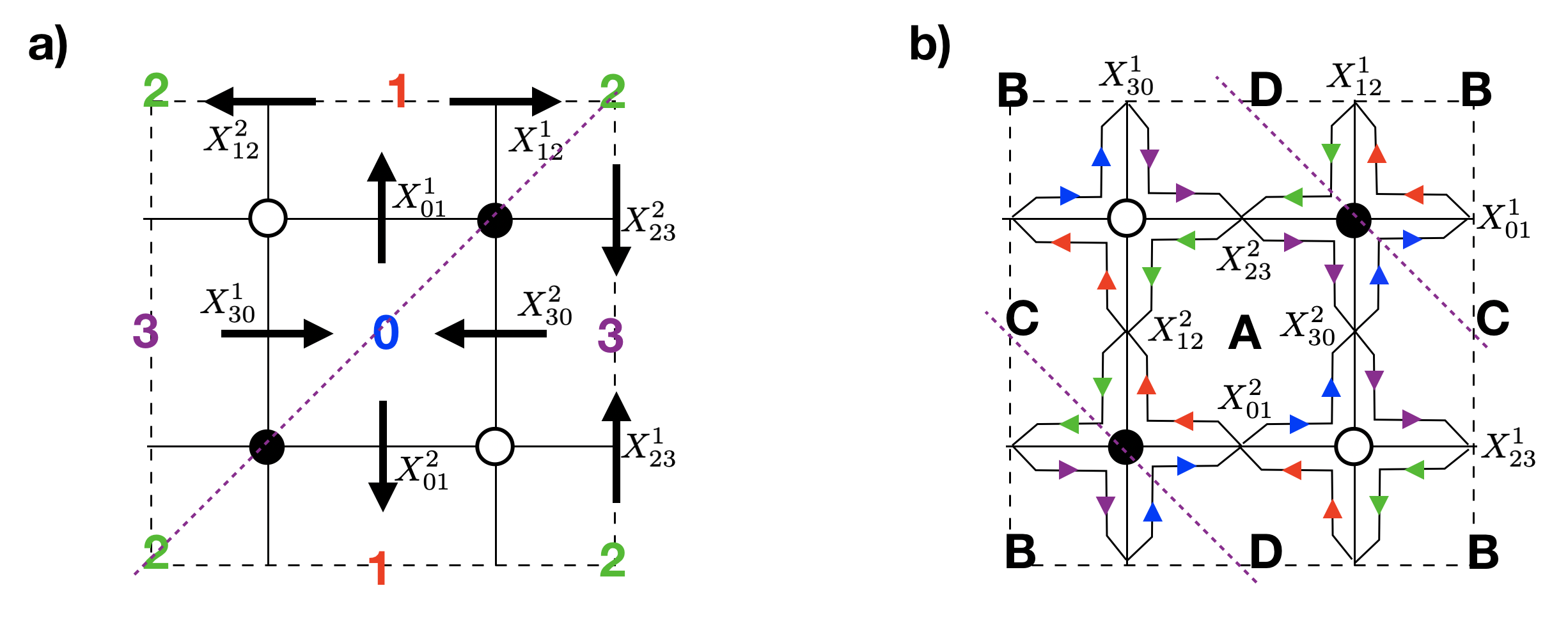}
\caption{\small a) The dimer diagram for the $\IF_0$ theory, with arrows indicating the  bifundamental chiral multiplets. We also display the $\IZ_2$ involution as a reflection with respect to the fixed line, indicated as a dashed violet line. b) The mirror Riemann surface with the D6-brane cycles in colors, and with the location of the bifundamental chiral multiplets at their intersections. The action of the $\IZ_2$ involution matching that in Figure (a) corresponds to an antiholomorphic action, of the kind studied in section \ref{sec:g2}.}
\label{fig:z2-action-f0}
\end{center}
\end{figure}

Let us quickly go through another example, the $\IF_0$ theory, obtained from D3-branes at the CY3 singularity given by the complex cone over $\IF_0$. This is equivalent to a $\IZ_2$ orbifold of the conifold $\{xy-zw=0\,|\,x,y,z,w\in\IC\}$, whose generator $\theta$ acts as
\beqa
\theta:(x,y,z,w)\to (-x,-y,-z,-w)\, .
\eeqa 
The geometry is toric, hence the theory has a dimer diagram description\footnote{As discussed in \cite{Feng:2001xr} the theory actually  has two toric phases, related by Seiberg duality \cite{Beasley:2001zp,Feng:2001bn}.} shown in Figures \ref{fig:mirror-f0}a, \ref{fig:z2-action-f0}a. The gauge theory is
\beqa
& SU(N)_0\times SU(N)_{1}\times SU(N)_2\times SU(N)_{3} &\nonumber \\
& 2 \,[\, (\fund_0,\antifund_1)+(\fund_1,\antifund_2)+(\fund_2,\antifund_3)+(\fund_3,\antifund_0)\,]\, .&
\label{f0}
\eeqa
There is also a quartic superpotential which can be read easily from the dimer diagram and which we skip. For completeness, we display in Figure \ref{fig:mirror-f0}b the mirror Riemann surface obtained via the untwisting procedure in \cite{Feng:2005gw}.

%%%%%%%%%%%
%%%%%%%%%%%

In Figure \ref{fig:z2-action-f0}a we have indicated the $\IZ_2$ involution required to build the 7d geometry $\IX_7$. Specifically, the gauge groups $SU(N)_1$ and $SU(N)_3$ are exchanged, while $SU(N)_0$ and $SU(N)_2$ are mapped to themselves. The action on the bifundamental matter is
\beqa
R:\,& X_{01}^1\leftrightarrow {\ov X}_{30}^2\quad ,\quad X_{12}^1 \leftrightarrow {\ov X}_{23}^2 & \nonumber \\
& X_{23}^1 \leftrightarrow {\ov X}_{12}^2 \quad ,\quad X_{30}^1\leftrightarrow {\ov X}_{01}^2\, .
&
\label{}
\eeqa 

%%%%%%%%%%%
%%%%%%%%%%%

In the mirror geometry this action corresponds to an antiholomorphic involution of the kind discussed in section \ref{sec:g2}, see Figure \ref{fig:z2-action-f0}b for its restriction to the mirror Riemann surface. The action on the fields allows to easily obtain the geometric action on the type IIB geometry. The coordinates of the parent conifold are constructed as the gauge invariant mesons
\beqa 
& x=X^1_{01}X^1_{12}X^1_{23}X^1_{30} \quad , \quad y=X^2_{01}X^2_{12}X^2_{23}X^2_{30}& \nonumber\\
&z=X^1_{01}X^2_{12}X^1_{23}X^2_{30} \quad , \quad w=X^2_{01}X^1_{12}X^2_{23}X^1_{30}\, ,&
\label{z2-fields-f0}
\eeqa
modulo F-term relations. Using (\ref{z2-fields-f0}), the action is
\beqa 
R\, :\, (x,y,z,w) \, \to\, ({\ov y},{\ov x},{\ov z},{\ov w})\, ,
\eeqa 
which again defines a 7d $G_2$ orbifold, preserving half of the supersymmetries.
Following the above Chiral Cone construction, we have a boundary configuration for a 4d $\NN=1$ chiral theory with spectrum (\ref{f0}) coupled to gravity.

\subsubsection{Deformation fractional branes}
\label{sec:fractional}

In the constructions we have encountered so far, the mechanism gapping the chiral non-anomalous set of fermions is not amenable to a simple field theoretical analysis, as it either involves transitions between tensor and hypermultiplets in 6d, or tachyon condensation processes. In this section we describe a large class of examples, based on the construction of 7d cones over D3-branes in (orbifolded) CY3 singularities, admitting a simple description in terms of supersymmetric gauge theory dynamics. In this sense, they provide a realization, in theories coupled to gravity, of the symmetric mass generation field theory mechanisms in  \cite{Razamat:2020kyf,Tong:2021phe}, also \cite{Wang:2022ucy} for a review.

The models are based on the use of fractional branes. In the previous discussion, we have considered regular D3-branes at the CY3 singularity $\IX_6$, which correspond (in the toric setup) to all gauge factors having equal rank. They describe systems where the D3-branes can move off the singular point into the CY3 bulk. The resulting gauge theories are 4d $\NN=1$ SCFT's, dual to AdS$_5\times\IY_5$ \cite{Klebanov:1998hh,Morrison:1998cs}. 

By fractional branes we mean general rank assignments constrained by anomaly cancellation\footnote{By this we mean that the anomalies are cancelled even when additional brane antibrane pairs are introduced. This is equivalent to requiring cancellation of the underlying RR tadpoles, which is in general a stronger condition than mere anomaly cancellation \cite{Uranga:2000xp}.}. The resulting gauge theories are no longer exactly conformal and, as explained in \cite{Franco:2005zu}, the different kinds of fractional branes can be classified according to their infrared behaviour: (1) $\NN=2$ fractional branes have exact Coulomb branches, describing the motion of D3-branes along a complex plane of singularities, and their gravity duals include enhan\c{c}on singularities \cite{Johnson:1999qt}; (2) Deformation fractional branes have strong infrared gauge dynamics generating a mass gap, and their gravity duals correspond to a complex deformation of the original singularity \cite{Franco:2005fd}, generalizing the mechanism in \cite{Klebanov:2000hb}; (3) DSB branes have strong infrared dynamics breaking supersymmetry \cite{Berenstein:2005xa,Franco:2005zu,Bertolini:2005di} and producing runaway behaviours \cite{Franco:2005zu,Intriligator:2005aw} \footnote{The introduction of orientifolds provides a further class fractional branes, which break supersymmetry with a presumably stable vacuum \cite{Franco:2007ii,Argurio:2020dkg,Argurio:2022vfq}; note that these have been conjectured in \cite{Buratti:2018onj} not to admit an AdS-like holographic dual due unstabilities upon the addition of regular D3-branes.}.

For our discussion, $\NN=2$ fractional branes tend to correspond to non-chiral theories, while DSB branes actually have no stable vacuum, hence we focus on deformation fractional branes. The most familiar realization corresponds to the fractional brane of the conifold theory \cite{Klebanov:2000hb}, but it is easy to use dimer techniques to generate large classes of examples \cite{Franco:2005fd}. 

\begin{figure}[htb]
\begin{center}
\includegraphics[scale=.35]{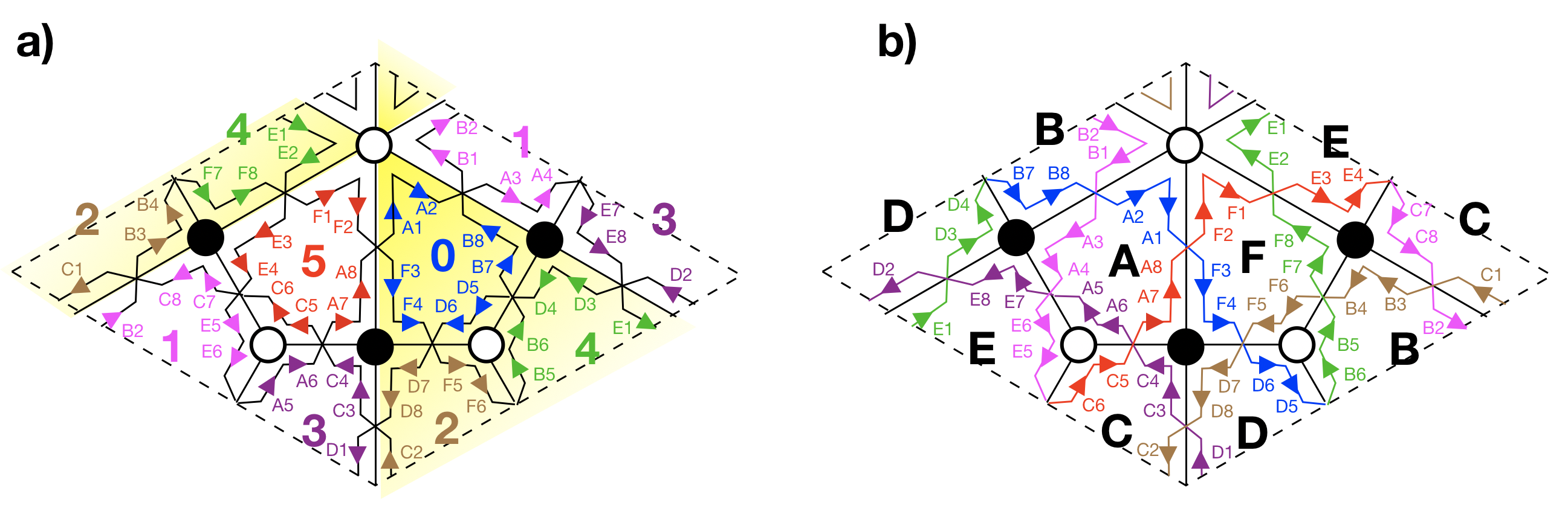}
\caption{\small a) The dimer diagram for the dP$_3$ theory, with its set of zig-zag paths. The yellow area corresponds to the set of faces covered by the deformation fractional branes. b) The mirror Riemann surface obtained via the untwisting procedure in \cite{Feng:2005gw}.}
\label{fig:mirror-dp3}
\end{center}
\end{figure}

\begin{figure}[htb]
\begin{center}
\includegraphics[scale=.3]{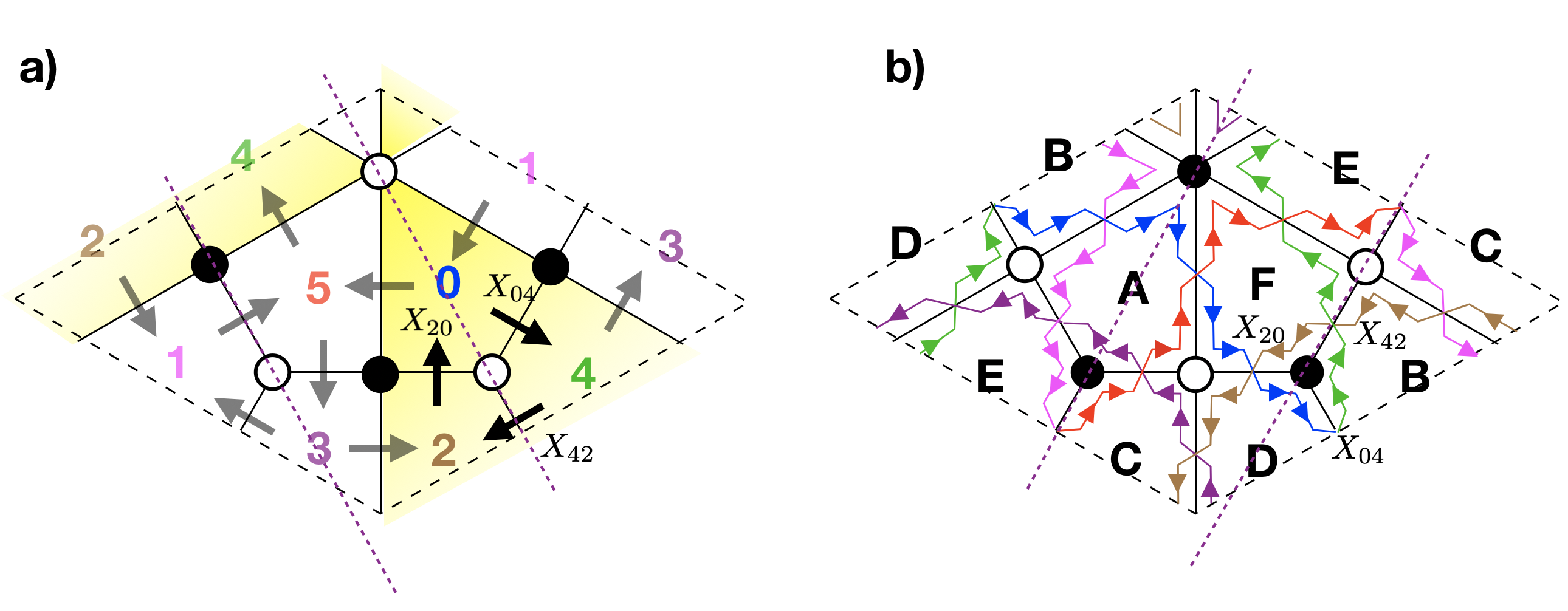}
\caption{\small a) The dimer diagram for the dP$_3$ theory, with arrows indicating the  bifundamental chiral multiplets. We have highlighted the faces and arrows present for the deformation fractional branes. We also display the $\IZ_2$ involution as a reflection with respect to the fixed line, indicated as a dashed violet line. b) The mirror Riemann surface with the D6-brane cycles in colors, and with the location of the bifundamental chiral multiplets at their intersections, and the antiholomorphic $\IZ_2$ action.}
\label{fig:z2-action-dp3}
\end{center}
\end{figure}

For concreteness, we focus on a deformation fractional brane in the complex cone over dP$_3$, see \cite{Franco:2005fd} for details. The dimer diagram if shown in Figures \ref{fig:mirror-dp3}a, \ref{fig:z2-action-dp3}a. For completeness, we display in Figure \ref{fig:mirror-dp3}b the mirror Riemann surface obtained via the untwisting procedure in \cite{Feng:2005gw}. 

As explained above, we allow for general ranks, compatible with anomaly cancellation. The most general solution for the rank vector is
\beqa
\vec{N}=N(1,1,1,1,1,1)+ M(1,0,1,0,1,0)+ P_1(1,0,0,1,0,0)+P_2(0,1,0,0,1,0)\, .\quad\quad
\eeqa
The set of $N$ corresponds to regular D3-branes, the sets of $P_1$ or $P_2$ corresponds to two deformations to conifold singularities, and the set of $M$ corresponds to a deformation directly to a smooth geometry. Although the construction of the orbifolded 7d cone can be carried out in more general cases, we focus on the simple choice $N=P_1=P_2=0$.
The gauge theory is given by
\beqa 
&SU(M)_0\times SU(M)_2\times SU(M)_4 &\nonumber \\
& (\fund_0,{\bf 1},\antifund_4)\,+\,({\bf 1},\antifund_2,\fund_4)\,+\, (\antifund_0,\fund_2,{\bf 1}) &\nonumber\\
&W=  X_{04}X_{42} X_{20}\, .&
\label{dp3}
\eeqa

%%%%%%%%%%%
%%%%%%%%%%%

In Figure \ref{fig:z2-action-dp3}a we have indicated the $\IZ_2$ involution required to build the 7d geometry $\IX_7$. Although the involution can be determined in general, we just specify the action on those fields corresponding to the deformation fractional branes. Specifically, the gauge group $SU(M)_0$ is mapped to itself, while $SU(M)_2$ and $SU(M)_4$ are exchanged. On the bifundamental matter, $X_{42}$ is invariant while $X_{04}$ and $X_{20}$ are exchanged. As in previous examples,  this action corresponds to an antiholomorphic involution in the mirror geometry.

%%%%%%%%%%%
%%%%%%%%%%%

Following the by now familiar Chiral Cone construction, we have a boundary configuration for the 4d $\NN=1$ chiral theory (\ref{dp3}) coupled to gravity. It is now easy to check the gauge field theory mechanism by which this gauge theory is gapped, as we quickly sketch below, see \cite{Franco:2005fd} for more detailed discussion. As the configuration approaches the boundary of spacetime i.e. the tip of the cone, the wrapped cycles are shrinking, hence the gauge factors are driven to strong coupling. Consider for simplicity that we consider one of the gauge factors, say $SU(M)_0$, to have a larger dynamical scale than the other two. Then, as we run to the infrared, we account for its strong gauge dynamics, taking $SU(M)_2$ and $SU(M)_4$ as global flavour symmetries. The $SU(M)_0$ theory has $N_f=N_c$, so that it confines and has a quantum deformed mesonic moduli space. Equivalently, we may replace it by the corresponding Seiberg dual, which has trivial gauge group, and a set of mesons $M_{42}$ in the $(\antifund_2,\fund_4)$. The gauge theory is thus
\beqa 
& SU(M)_2\times SU(M)_4 &\nonumber \\
& (\antifund_2,\fund_4)\,+\,(\fund_2,\antifund_4)&\nonumber\\
&W=  M_{42}X_{24}\, . &
\eeqa 
Namely, the composite mesons transform in the conjugate representation of the (spectator) bifundamental $X_{24}$, and in fact they get a mass term via the superpotential coupling. Hence the original chiral non-anomalous set of fermions is gapped, in the spirit of symmetric mass generation. In the language of the original gauge theory (\ref{dp3}), the boundary condition at the ETW brane relates the bifundamental $X_{24}$ with the value of a composite degree of freedom of $X_{40}$ and $X_{02}$. Similar statements hold if we consider some other of the original gauge factors to confine first; actually, due to the cyclic symmetry of the theory, we expect that all gauge factors confine at the same scale and there is a combination of the above phenomenon taking place for each gauge factor. The whole gauge theory analysis is efficiently encoded in simple manipulations in the dimer diagram, as shown in \cite{Franco:2005fd,GarciaEtxebarria:2006aq}.

In the context of holography, the above kind of behaviour was argued in \cite{Franco:2005fd} to occur for any deformation fractional brane. Namely, the set of gauge factors in the fractional D3-brane gauge theory has a subsector that confines and develops a quantum deformed mesonic moduli space, resulting in a gapping of the remaining theory. In the gravity dual, this is described by a generalization of \cite{Klebanov:2000hb}, in which the quantum deformed moduli space describes the complex deformation of $\IX_6$, by replacing its singular tip by a 3-cycle whose size is fixed by 3-form fluxes dual to the number of fractional brane number $M$. 

There exist systematic techniques to build gauge theories with fractional deformation branes and study the complex deformations in the corresponding toric geometries \cite{GarciaEtxebarria:2006aq}. Hence, it is straightforward to use these tools to construct further explicit examples. We hope that the above discussion suffice to illustrate the main points, and refrain from further discussions.

\subsection{The dynamical cobordism}
\label{sec:dyn-cob-d3s}

In this section we describe explicitly the Cone Constructions as solutions of the theory after compactification on the base of the cone, and show that they correspond to dynamical cobordisms in the precise sense of \cite{Angius:2022aeq}. Namely, the solutions describe a running scalar which attains infinite field space distance at a curvature singularity at finite spacetime distance, at which spacetime ends.

The discussion holds for the general class of cone constructions, in particular also the 6d examples in section \ref{sec:transitions}. Nevertheless, for the sake of concreteness we focus on the particular class of 4d theories arising from D3-branes at general CY3 singularities, in sections \ref{sec:cone-singus} and \ref{sec:z2-quotients}. We moreover treat the D3-branes as probes, hence the solution is mainly associated to the compactification over the 6d geometry $\IY_6$ given by the suspension of the base $\IY_5$ of the CY3 cone $\IX_6$.

\subsubsection{Dimensional reduction from $D$ to $n$ dimensions}
\label{sec:reduction}

We start with the $D-$dimensional Einstein-Hilbert action:
 \begin{equation}
     S_D = \frac{1}{2} \int d^Dx \sqrt{-G_D} R_D \, ,
     \label{action_S_D}
 \end{equation}
and consider the following ansatz for compactification on a $p$-dimensional space $\IY_p$ parametrized by $y^i$, $i=n+1,\ldots, n+p=D$ :
 \begin{equation}
     ds^2_D = e^{2 \alpha \rho(x)} g_{\mu \nu} dx^{\mu} dx^{\nu} + e^{2 \beta \rho(x)} g_{ij} d y^i d y^j\, ,
     \label{metric_ansatz}
 \end{equation}
where $\alpha$ and $\beta$ are constant coefficients. The relation among the $D$-, $n-$ and $p$-dimensional curvature scalars and the scalar $\rho$ is
\begin{equation}
\begin{split}
    R_D & =   e^{-2 \alpha \rho} \left\lbrace R_n + R_p e^{2 (\alpha - \beta) \rho} -2 \left[ (\alpha - \beta)(n-1) + \beta (D-1) \right] \Delta \rho - \right.  \\
    & \left. - \left[ (\alpha - \beta)^2 (n-2)(n-1) + 2 \beta ( \alpha - \beta) (n-2)(D-1) + \beta^2 (D-1)(D-2) \right] \vert d \rho \vert^2 \right\rbrace. \\
\end{split}
\end{equation}
Using this and the relation between determinants of the metrics, the action $S_D$ (\ref{action_S_D}) becomes:
\beqa
     S_D = \frac{1}{2} \int_{{\IY}_p}  \int_{\mathcal{M}_n} d^p y \,d^n x \,\sqrt{-g_n} \,\sqrt{g_p} \, e^{\left[ (n-2) \alpha +p \beta \right] \rho} \left\lbrace  R_n + R_p \,e^{2(\alpha - \beta)\, \rho}   - C_1 \Delta \rho 
    - C_2 \vert d \rho \vert^2  \right\rbrace \, ,\nonumber
\label{action_D_dim}
\eeqa
with 
\beqa
C_1&=&2 \left[ (\alpha - \beta)(n-1) + \beta (n+p-1) \right] \\
C_2&=& (\alpha - \beta)^2 (n-2)(n-1) +2 \beta (\alpha - \beta)(n-2)(n+p-1) + \beta^2 (n+p-2)(n+p-1)\, , \nonumber
\eeqa
Redefining to the Einstein frame (we exclude $n\neq 2$ in what follows) and to canonically normalized scalar kinetic term with the conditions
\beqa
    \beta = - \frac{n-2}{p} \alpha\quad ,\quad  \alpha^2 = \frac{p}{(n-2)(n+p-2)}\, ,
\label{alpha-beta}
\eeqa   
and integrating by parts the Beltrami operator term, we obtain
\begin{equation}
    S_D = \frac{1}{2} \int_{\IY_p} \int_{\mathcal{M}_n} d^p y\, d^nx \sqrt{-g_n} \sqrt{g_p}  \left\lbrace  R_n + R_p \,e^{2 \frac{p+n-2}{p} \alpha \rho}  - \vert d \rho \vert^2  \right\rbrace\, .
    \label{action_D_dim_simplified}
\end{equation}

\subsubsection{Compactification on the base of the cone}
\label{sec:base}

At this point we specialize to the setup of 4d solutions in type IIB string theory, hence we set $D=10$ and $p=6$, so the action \eqref{action_D_dim_simplified} becomes:
\begin{equation}
    S_{10} = \frac{1}{2} \int_{\IY_6} \int_{\mathcal{M}_4} d^6 y\, d^4x\, \sqrt{-g_4} \sqrt{g_6} \left\lbrace  R_4 + R_6 \,e^{\pm 2 \sqrt{2/3} \rho} - \vert d \rho \vert^2 \right\rbrace\, .
    \label{action_10d_simplified}
\end{equation}
where the two signs correspond to the two solutions for $\alpha$ in (\ref{alpha-beta}).

For constant curvature compact space, for instance $\IS^6/\IZ_k$, arising in singularities from orbifolds of flat space (e.g. the dP$_0$ example in sections \ref{sec:dp0-example}, \ref{sec:examples}), it is straightforward to integrate over the internal space and get
\begin{equation}
\begin{split}
    S_4 \propto \frac{{\cal V}_{\IS^6}}{2k} \int d^4x \sqrt{-g_4} \left\lbrace R_4 - \vert d \rho \vert^2  + \frac{30}{R_0^2} e^{\pm 2 \sqrt{2/3} \rho}  \right\rbrace.\\
\end{split}
    \label{reduced_action_4d_over_S6}
\end{equation}
where $R_0$ is the radius of the covering $\IS^6$, and ${\cal V}_{\IS^6}=\frac{16}{15} \pi^3 R_0^6$ is its volume.  
More in general, we  want to consider the configurations in sections \ref{sec:examples-singus}, \ref{sec:z2-quotients}, so we consider compactification on a 6d geometry $\IY_6$ (eventually the base of a 7d cone) given by a suspension of a 5d geometry $\IY_5$ (eventually the base of the 6d CY3 cone) over the segment $\theta \in \left[ - \pi /2 , \pi /2 \right]$, namely recalling (\ref{cone-7d-two}), (\ref{cone-7d}),
\beqa
ds_{\IY_6}^2=d\theta^2+\cos^2\theta \,ds_{\IY_5}^2\, .
\eeqa
Hence we may express the action in terms of the geometric properties of $\IY_5$. In particular, using the relation of curvatures
\begin{equation}
    R_{\mathbf{Y_6}}= \frac{1}{\cos^2 \theta} \left[ R_{\mathbf{Y_5}} -10 \cos^2 \theta -20 \sin^2 \theta \right]\, ,
\end{equation}
and integrating over $\theta$ in \eqref{action_10d_simplified} we obtain\footnote{The expressions correspond to the simple compactification of section \ref{sec:examples-singus}. The $\IZ_2$ quotient for models of section \ref{sec:z2-quotients} lead to additional simple factors of 2.}: 
\begin{equation}
\begin{split}
    S_{10} = 
    & 
    \frac{8}{15} \int_{\mathbf{Y_5}} \int_{\mathcal{M}_4} d^5 y d^4x \sqrt{-g_4} \sqrt{g_{\mathbf{Y_5}}} \left\lbrace R_4 - \vert d \rho \vert^2  + \frac{5}{4} \left(-12 + R_{\mathbf{Y_5}} \right) e^{\pm 2\sqrt{2/3} \rho}\right\rbrace. \\
    =& \frac{8}{15} Vol_{\mathbf{Y_5}} \int_{\mathcal{M}_4} d^4x \sqrt{-g_4} \left[ R_4 - \vert d \rho \vert^2 -15 e^{\pm 2\sqrt{2/3} \rho}\right] + \frac{2}{3} \cdot A \int_{\mathcal{M}_4} d^4x \sqrt{-g_4} e^{\pm 2 \sqrt{2/3} \rho} \, ,
    \label{the-action}
\end{split}
\end{equation}
where $A= \int_{\mathbf{Y_5}} d^5 y \sqrt{g_{\mathbf{Y_5}}} R_{\mathbf{Y_5}}$.

\subsubsection{Dynamical cobordism and scaling relations}
\label{sec:scaling}

We can now match the 10d metric with the compactification ansatz \eqref{metric_ansatz}: 
\begin{equation}
    ds_{10}^2=e^{2\alpha\rho}ds_4^2+e^{2\beta\rho}R_0^2ds_6^2=\left(dx^0\right)^2 + \left( dx^1 \right)^2 + \left( dx^2 \right)^2 + dr^2+r^2ds_{\bf{Y}_6}^2,
\end{equation}
where $R_0$ is a reference value for the internal space size. 

The matching of the internal part leads to the profile of the breathing mode $\rho(r)$
   \begin{equation}
       R_0^2 e^{2 \beta \rho} ds_{6}^2= r^2 ds^2_{\bf{Y}_6} \quad \Longrightarrow\quad
       \rho(r)= \frac{1}{\beta} \log\left(\frac{r}{R_0}\right).
       \label{radion_profile}
   \end{equation}
  Comparing the non-compact directions, we can extract the 4d metric. Using the above scalar profile and the Einstein frame condition, we obtain
      \begin{equation}
       ds_4^2 = e^{6 \log \left( r/R_0 \right)} \left[ (dx^1)^2+(dx^2)^2+(dx^3)^2+dr^2\right]. 
       \label{4d_metric_r}
   \end{equation}
   This can be recast in the standard local dynamical cobordism form in \cite{Angius:2022aeq}, as follows. We redefine coordinates to encode the physical spacetime distance, via 
   \begin{equation}
      dy^2 = \left(\frac{r}{R_0}\right)^6dr^2\quad \Longrightarrow\quad
       y=\int^{r}_0\left(\frac{\tilde{r}}{R_0}\right)^3d \tilde{r} = \frac{r^4}{4R_0^3}\sim r^4\, .\quad\quad
   \end{equation}
In terms of the new coordinate the metric \eqref{4d_metric_r} takes the form:
\begin{equation}
    ds^2_4 = e^{\frac{3}{2}\log(4y/{R_0})}\left[ (dx^1)^2+(dx^2)^2+(dx^3)^2\right]+dy^2.
\end{equation}
The metric and scalar profiles obey the local dynamical cobordism ansatz in \cite{Angius:2022aeq}
\beqa
 ds^2=e^{-2\sigma(y)}ds_{n-1}^2+dy^2 \quad ,\quad \sigma(y)\simeq \frac 2{\delta^2}\log y \quad ,\quad
\rho(y)\simeq -\frac 2\delta\log y\, ,
\eeqa
with a critical exponent given by
\beqa 
\delta= \frac{2}{3} \sqrt{6}\, ,
\eeqa
Note that this agrees with the scaling of the potential in (\ref{the-action}) as $V\sim \exp (\delta\phi)$ \cite{Angius:2022aeq}. As also shown in this reference, the quantity $\delta$ also controls the scalings of the field space distance ${\cal D}$ and the curvature $R$ with the spacetime distance $\Delta$ via
\beqa
\Delta\sim e^{-\frac\delta{2}\cal D}\quad ,\quad |R|\sim e^{-\delta\cal D}\, .
\eeqa
This result confirms that our cones constructions can be regarded as dynamical cobordism solutions of the theory obtained upon compactification on the base of the cone. We emphasize that similar computations lead to this conclusion also for other setups, such as the 6d examples in section \ref{sec:transitions}, or the alternative 4d setup in appendix \ref{app:g2-cone-example}.

\section{Conclusions}
\label{sec:conclu}

In this paper we have studied the construction of boundary configurations for several large classes of 6d and 4d chiral theories arising from string theory, hence including gravity. The boundary configurations are mostly constructed using cones over codimension 1 slices at which some interesting physics takes place, such as chirality changing phase transitions associated to the ending of some lower-dimensional brane with chiral worldvolume theory, or to some $\IZ_2$ quotient leading to reduced but non-trivial supersymmetry at the tip of the cone. 

Interestingly, the physical mechanisms associated to the boundary configuration often admit a field theory interpretation, albeit at strong coupling. For example, in the 6d cases it is associated to the transitions trading one 6d $\NN=1$ tensor multiplet for 29 hypermultiplets, which does not admit a lagrangian description, while in the 4d case they are often related to confinement and pairing up of fundamental chiral multiplets with composite mesons, as in the case of deformation fractional branes.

This work is a useful stepping stone in the general program of building boundary configurations for general chiral theories coupled to gravity. Clearly, there remain many important challenges in this plan, for instance:

$\bullet$ Most prominently, an open question is the definition of boundary configurations for the chiral 10d string theories. One interesting possibility is to exploit their realization as the endpoint of closed string tachyon condensation of higher-dimensional supercritical strings \cite{Hellerman:2004qa,Hellerman:2006nx,Hellerman:2006ff,Hellerman:2006hf,Hellerman:2007fc,Hellerman:2010dv,Garcia-Etxebarria:2014txa}, to allow for some version of the cone construction described in our examples.

$\bullet$  In the 6d setup we have found boundary configurations involving chirality changing phase transitions. It would be interesting to build explicit examples of boundary configurations for 4d chiral theories from 4d chirality changing phase transitions as well.

$\bullet$ In the context of 4d dimensional examples, we have managed to provide boundary configurations for bulk chiral theories with additional $\IZ_2$ symmetries, either under exchange of whole gauge sectors, or as involutions of a given quiver gauge theories. It would be interesting to explore boundary conditions for general 4d theories, possibly not enjoying such symmetries.

$\bullet$ The connection of our cone constructions with those involved in the derivation of SymTFTs, discussed in section \ref{sec:symtft} seems to provide a new interesting tool to analyze the topological properties of chirality changing phase transitions. One may hope to use these tools to gain a better understanding of the basic (yet highly non-trivial) 6d transition turning one tensor multiplet into 29 hypermultiplets.

We hope to come back to these and other questions in the coming future.

%\newpage

\section*{Acknowledgments}

We are pleased to thank Matilda Delgado, Bjoern Friedrich, I\~naki Garc\'\i a-Etxebarria, Arthur Hebecker, Jes\'us Huertas, Luis Ib\'a\~nez, Fernando Marchesano, Miguel Montero, Irene Valenzuela, Johannes Walcher and Xingyang Yu for useful discussions. R.A. thanks the CERN Theory Physics Department for hospitality during the completion of this work and the ERC Starting Grant QGuide-101042568 - StG 2021 for supporting this stay. C. W. thanks the hospitality of the Department of Mathematical Sciences of Durham University, where part of this work was carried out.
This work is supported through the grants CEX2020-001007-S, PID2021-123017NB-I00 and  ATR2023-145703 funded by MCIN/AEI/10.13039/501100011033 and by ERDF A way of making Europe. 

\newpage

\appendix

\section{More systems of intersecting D6-branes with boundaries}
\label{app:more-intersecting}

In this appendix we consider configurations of intersecting semi-infinite D6-branes with boundaries defined by NS5-branes, generalizing those in section \ref{sec:open-intersecting}. We will similarly find that, due to the presence of additional emitted D4-branes, the cone construction leads to dynamical cobordisms for theories which are ultimately non-chiral. These extra examples thus confirm that this class of construction is not optimal to obtain boundary configurations for chiral theories.

We would like to consider a configuration similar to that in section \ref{sec:open-intersecting}, but with two semi-infinite D6-brane stacks, each ending on an NS5-brane. Because the number of D6-branes emitted by an NS5-brane is determined by the Romans mass $m$, we have to consider the two stacks to have the same number of D6-branes $N_1=N_2=m$. Actually, this example is part of a slightly more general class, in which we consider stacks of half D6-branes on both sides of each NS5-branes, with the numbers differing by $m$ units consistently with the Freed-Witten effect on the NS5-branes. Namely, we take a set of D6-branes along 0123 456, split in two semi-infinite stacks of half-D6-branes by an NS5-brane (dubbed NS5$_1$) at 012 456 at $x^3=0$, so we have $N_1$ D6$_1$- and $N_1'$ D6$_1'$-branes at $x^3>0$ and $x^3<0$ respectively, with $N_1-N_1'=m$. Similarly, we introduce another set of D6-branes along 0123 789, split in two semi-infinite stacks of half-D6-branes by an NS5-brane (dubbed NS5$_2$) at 012 789 at $x^3=0$, so we have $N_2$ D6$_2$- and $N_2'$ D6$_2'$-branes at $x^3>0$ and $x^3<0$ respectively, with $N_2-N_2'=m$.

\begin{figure}[htb]
\begin{center}
\includegraphics[scale=.3]{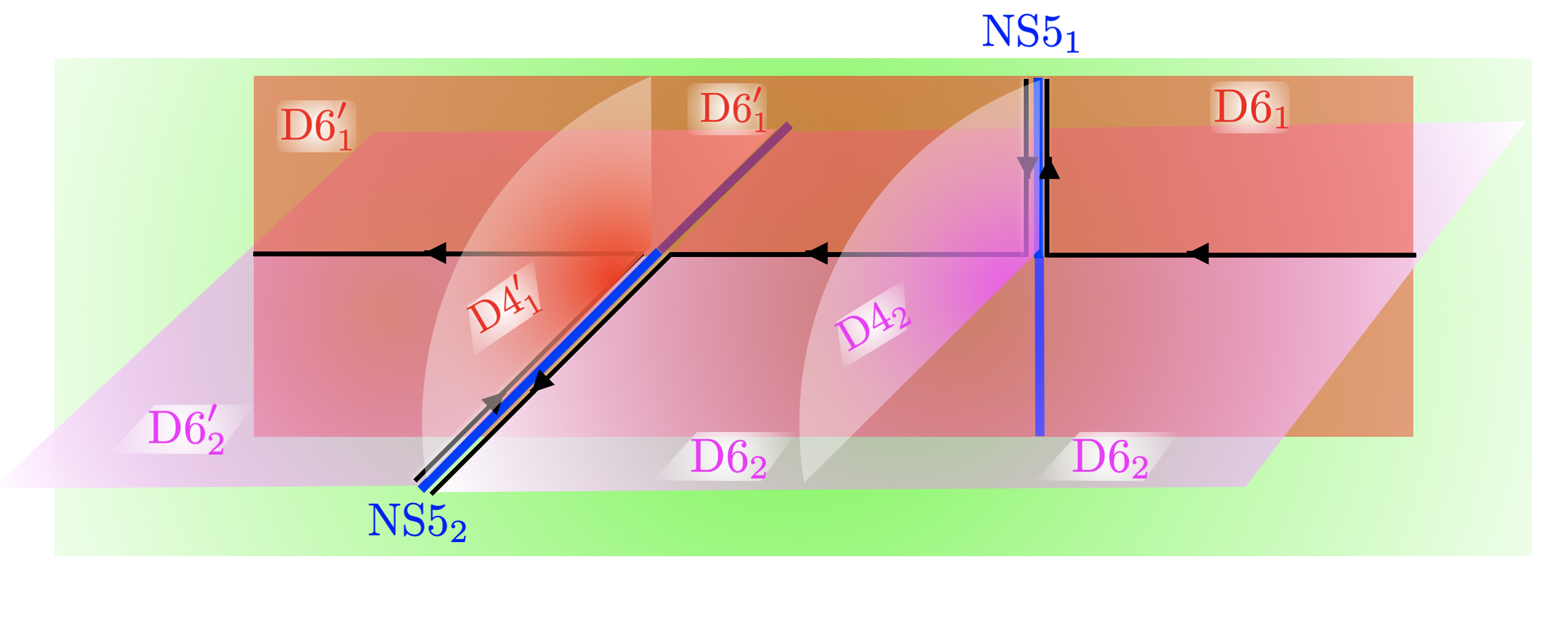}
\caption{\small Stacks of semi-infinite D6-branes separated by (slightly separated) NS5-branes, and intersecting over 4d.  We have depicted the D4$_1'$-branes stretched between the D6$_1'$-branes and the NS5$_2$-brane, and the D4$_2$-branes stretched between the D6$_2$-branes and the NS5$_1$-brane. We have indicated with black lines and arrows the location of 4d chiral fermions in the configuration. They complete continuous paths, displaying the consistency with anomaly inflow from the bulk of the D6/D4-branes.}
\label{fig:separate}
\end{center}
\end{figure}

Configuration of intersecting NS5-branes are often very non-trivial (see e.g. \cite{Hanany:1996hq}), hence we will thus regulate our setup by taking one of the NS5-branes (e.g. the NS5$_1$) at a nonzero value\footnote{The situation with $\epsilon<0$ can be studied similarly, and leads to slightly different intermediate spectra. This indicates that the limit $\epsilon\to 0$ is presumably not smooth, signalling a possibly non-lagrangian strongly coupled theory for the coincident NS5-brane case.} of $x^3=\epsilon>0$, see Figure \ref{fig:separate}. 

We now have four sets of gauge fields on the D6$_1$-, D6$_1'$-, D6$_2$- and D6$_2'$-branes. Also, we have a 4d intersection of the D6$_1$- and D6$_2$-branes on the half-space along 012 and at $x^3>\epsilon$, giving a 4d chiral fermion in the bifundamental $(\fund_1,{\bf 1};\antifund_2,{\bf 1})$, which ends at the NS5$_1$-brane at $x^3=\epsilon$, an intersection of the D6$_1'$- and D6$_2$- branes on the space along 0123 and the segment $0<x^3<\epsilon$, giving a 4d chiral fermion in the $({\bf 1},\fund_{1}';\antifund_2,{\bf 1})$, which ends at the NS5$_2$-brane at $x^3=0$, and a 4d intersection of the D6$_1'$- and D6$_2'$-branes on the half-space along 012 and at $x^3<0$, giving a 4d chiral fermion in the bifundamental $({\bf 1},\fund_1';{\bf 1},\antifund_2')$. 

As in section \ref{sec:open-intersecting}, the discontinuity of the chiral fermion spectrum and the apparent mismatch of anomalies across the NS5-branes indicates that the configuration is missing extra ingredients. These are again given by emitted D4-branes stretching between the NS5- and the D6-branes, as follows. Using by now familiar arguments, there are D4-branes (dubbed D4$_2$-branes) attached to the D6$_2$-branes and ending on the NS5$_1$-brane, and D4$_1'$-branes attached to the D6$_1'$-branes and ending on the NS5$_2$-brane. As explained, we should think about the D4-branes as carrying the same worldvolume gauge group as the D6-branes to which they are attached (hence the similar notation). The final configuration is depicted in Figure \ref{fig:separate}, and shows the extra D4-branes guarantee the continuity of the 4d chiral fermion content, indicated by black lines with arrows.

%%%%%%%%%%%
%%%%%%%%%%%

It is now easy to consider the limit $\epsilon\to 0$ (or very small, for that matter) and to carry out the cone construction on this configuration, to obtain a dynamical cobordism of the 4d theory upon compactification on $\IS^6$. It is clear that, as in the example in section \ref{sec:open-intersecting}, the continuity of the lines supporting the 4d fermions implies that, in the cone perspective, for each line of incoming chiral fermions there is an outgoing line of fermions with the same quantum numbers. Hence there is a doubling of the spectrum of the 4d theory making it non-chiral.

We hope this example suffices to illustrate this is a general pattern for this class of configurations, as mentioned in the main text.

\section{Chiral Cone Constructions with Intersecting D6-branes}
\label{app:g2-cone-example}

In this appendix we describe an example of a 7d $G_2$ holonomy space $\IX_7$ given by a cone over a base $\IY_6$ with D6-branes wrapped on non-compact associative 4-cycles intersecting at points over $\IY_6$, and leading to a 4d chiral non-anomalous spectrum. The example is based on geometries constructed in \cite{Bryant:1989} and considered  in the holographic context in \cite{Acharya:2003ii} (see also \cite{Acharya:2005ez}). In our context, the system provides an explicit example realizing the ideas in section \ref{sec:g2} (albeit for a genuine, rather than barely, $G_2$ holonomy cone). We start with a review of the geometry, referring the reader to these works for further details.

The 7d space $\IX_7$ is a cone over a 6d base $\IY_6$ given by the coset $SU(2)^3/SU(2)$, where the quotient is by the diagonal subgroup. The space $\IY_6$ is topologically $\IS^3\times\IS^3$, and the metric is induced from the round metric in the parent $SU(2)^3$ space, namely
\beqa 
ds^2=dr^2+\frac{r^2}9({\omega_a}^2+{\tilde\omega}_a^2-\omega_a{\tilde\omega_a})\, ,
\label{bs-g2metric}
\eeqa
where $\omega_a$, ${\tilde\omega}_a$, $a=1,2,3$, are left invariant 1-forms of the two $\IS^3$'s, when regarded as $SU(2)$ groups. The metric (\ref{bs-g2metric}) has $G_2$ holonomy \cite{Bryant:1989}. The space $\IY_6$ admits an almost complex structure defined in terms of the complex frames
\beqa 
\eta_a=\omega_a+e^{2\pi i/3}{\tilde\omega_a}\, ,
\eeqa 
in terms of which the metric reads $ds^2=dr^2+r^2\eta_a{\ov\eta}_a$. One can define the (non-closed) K\"ahler and holomorphic forms
\beqa
\omega\equiv \frac i2 \eta_a\wedge{\ov\eta}_a &\quad , \quad & d\omega= -3{\rm Im}\, \Omega\nonumber \\
\Omega\equiv \eta_1\wedge\eta_2\wedge\eta_3 & \quad ,\quad & d{\rm Re}\, \Omega=-2\omega\wedge\omega\, .
\eeqa
The associative 3-form and coassociative 4-forms in $\IX_7$ are given in terms of these by
\beqa
\varphi=r^2dr\wedge \omega-r^3{\rm Im}\, \Omega\quad ,\quad *\varphi=r^3{\rm Re}\,\Omega\wedge dr+ \frac{r^4}2\omega\wedge\omega\, .
\label{asso-coasso}
\eeqa 
The metric is invariant under the order 6 permutation group of the 3 $\IS^3$'s in the parent space. This is generated by the order 2 exchange $\alpha$ of the two $\IS^3$'s in $\IY_6$, acting as $\alpha:\omega_a\leftrightarrow {\tilde\omega}_a$, and the order 3 cyclic permutation $\beta$ of the three $\IS^3$'s in the parent space, acting as $\beta:\omega_a\to -{\tilde\omega}_a\, , \; {\tilde\omega}_a\to \omega_a-{\tilde\omega_a}$. Equivalently, the actions are $\alpha:\eta_a\to \tau{\ov\eta}_a$ and $\beta:\eta_a\to \tau\eta_a$. The action over the forms (\ref{asso-coasso}) is $\alpha:\varphi\to-\varphi$, leaving $*\varphi$ invariant, while $\beta$ leaves both $\varphi$ and $*\varphi$ invariant. These actions can be extended to actions on the full cone $\IX_7$, for which, with a slight abuse of notation, we use the same names.

These symmetries are extremely useful to build supersymmetric cycles. As is well-known, it is notoriously difficult to construct calibrated submanifolds in $G_2$ manifolds, but, in analogy with the (similarly difficult construction of special lagrangian 3-cycles in CY3 spaces) particular examples can be obtained as the fixed point set of certain $\IZ_2$ involutions. In particular, one can build supersymmetric 4-cycles in $G_2$ manifolds as the fixed point set under a $\IZ_2$ action which flips the sign of $\varphi$ and leaves $*\varphi$ invariant. This is precisely the way $\alpha$ acts on $\IX_7$, hence the fixed point set of $\alpha$ in $\IX_7$ provides a supersymmetric 4-cycle in the cone geometry. Clearly, in the base $\IY_6$, it corresponds to the diagonal $\IS^3$ in the $\IS^3\times\IS^3$, hence in the full cone $\IX_7$ we get a cone over that $\IS^3$. Moreover, we can obtain other 4-cycles as images of the previous one under the action of $\beta$; or equivalently, as the fixed point set of the actions $\beta\alpha\beta^{-1}$ and $\beta^2\alpha\beta^{-2}$. Clearly, by symmetry, they are cones over $\IS^3$'s, obtained as diagonal combinations of consecutive $\IS^3$'s in the parent $SU(2)^3$ space.

Let us denote these 3-cycles in $\IY_6$ as $Q_i$, $i=0,1,2$. In the symplectic basis of 3-homology $[A]$, $[B]$ of $\IY_6=\IS^3\times\IS^3$, their homology classes can be expressed as
\beqa
[Q_0]=[A]+[B]\quad ,\quad [Q_1]=-[A]\quad ,\quad [Q_2]=-[B]\, .
\eeqa
This means that it is possible to wrap an equal number $N$ of D6-branes on each of these 3-cycles on $\IY_6$ in a way compatible with RR tadpole cancellation, since the total homology class vanishes. In the full 7d geometry, the D6-branes stretch also in the radial direction, namely they span the directions 012 and the 4-cycles given by the cones over $Q_i$.

We can now regard this configuration in the spirit of the cone construction, as a 4d theory obtained upon compactifying 10d type IIA on $\IY_6$ with the wrapped D6-branes, namely an intersecting D6-brane brane model \cite{Blumenhagen:2000wh,Aldazabal:2000dg, Aldazabal:2000cn} (see \cite{Ibanez:2012zz} for a review). The compactification space is not Calabi-Yau, but the determination of the spectrum is topological. Noticing that the intersection numbers of the 3-cycles are $=[Q_0]\cdot[Q_1]=[Q_1]\cdot[Q_2]=[Q_2]\cdot[Q_0]=1$ (and zero for those others not fixed by antisymmetry of the intersection product), we obtain a 4d gauge group, matter content, and superpotential
\beqa
&SU(N)^3& \nonumber\\
&(\fund,\antifund,{\bf 1})+({\bf 1},\fund,\antifund)+(\antifund,{\bf 1},\fund)&\nonumber\\
&W\sim X_{01}X_{12}X_{20}\, ,&
\label{gauge-g2-cone}
\eeqa
where the superpotential can be shown to arise from worldsheet instantons, and has a trace over color indices that we leave implicit. Note that we have removed the $U(1)$ factors, as the diagonal combination simply decouples, and the two other combinations are made massive by St\"uckelberg couplings.

In the spirit of the Chiral Cone construction, the full configuration in the cone $\IX_7$ describes a running solution of this theory, in which the scalar describing the size of $\IY_6$ runs along the radial coordinate of the cone, and diverges at finite distance in spacetime, corresponding to the shrinking of $\IY_6$ at the tip of the cone. Hence, it fits the picture of a dynamical cobordism \cite{Buratti:2021fiv,Angius:2022aeq,Blumenhagen:2022mqw,Blumenhagen:2023abk}, as emphasized in the main text and shown in section \ref{sec:dyn-cob-d3s}, hence providing a boundary configuration for the chiral theory.

Incidentally, we note that the gauge theory (\ref{gauge-g2-cone}) is the same as the theory (\ref{dp3}) realized on deformation branes of the complex cone over $dP_3$, despite being a completely different context. This allows us to borrow the discussion at the end of section \ref{sec:fractional} regarding the field theory analysis explaining how the chiral non-anomalous theory gets gapped. As in that discussion, the strong coupling gauge dynamics nicely dovetails the fact that the wrapped 3-cycles are shrinking as one approaches the boundary of spacetime, driving the gauge factors to strong coupling.

Let us finish with the observation in \cite{Acharya:2003ii} that the type IIA configuration admits a lift to M-theory, in which the D6-branes are fully geometrized (locally as $\IC^2/\IZ_N$ singularities). The final configuration is M-theory on an 8d $Spin(7)$ holonomy cone, and the gauge theory discussed above arises from the structure of codimension-4 singularities and the enhancements at the intersections of the corresponding singular loci. 

\bibliographystyle{JHEP}
\bibliography{mybib}

\end{document}